\documentclass[fleqn,usenatbib]{mnras}

\usepackage{newtxtext,newtxmath}
\usepackage{float}

\usepackage[T1]{fontenc}

\DeclareRobustCommand{\VAN}[3]{#2}
\let\VANthebibliography\thebibliography
\def\thebibliography{\DeclareRobustCommand{\VAN}[3]{##3}\VANthebibliography}

\usepackage{graphicx}	
\usepackage{amsmath}	

\usepackage{ae,aecompl}
\usepackage{relsize}
\usepackage{upgreek}
\usepackage{bm}
\usepackage{ulem}

\graphicspath{{./}{figures/}}

\title[Real Nature of Massive Clumps]{Physical Properties and Real Nature of Massive Clumps in the Galaxy}

\author[Z.-J. Lu et al.]{Zu-Jia Lu,$^{1}$\thanks{E-mail:luzujia@icc.ub.edu}
Veli-Matti Pelkonen,$^{1}$
Mika Juvela,$^{2}$
Paolo Padoan,$^{1,3}$
Troels Haugb{\o}lle$^{4}$
and \AA ke Nordlund$^{4}$
\\
$^{1}$Institut de Ci\`{e}ncies del Cosmos, Universitat de Barcelona, IEEC-UB, Mart\'{i} i Franqu\`{e}s 1, E08028 Barcelona, Spain\\
$^{2}$Department of Physics, PO Box 64, University of Helsinki, 00014, Helsinki, Finland\\
$^{3}$ICREA, Pg. Llu\'{i}s Companys 23, 08010 Barcelona, Spain\\
$^{4}$Niels Bohr Institute, University of Copenhagen, {\O}ster Voldgade 5-7, DK-1350 Copenhagen K, Denmark
}

\date{Accepted XXX. Received YYY; in original form ZZZ}

\pubyear{2021}

\begin{document}
\label{firstpage}
\pagerange{\pageref{firstpage}--\pageref{lastpage}}
\maketitle


\begin{abstract}

Systematic surveys of massive clumps have been carried out to study the conditions leading to the formation of massive stars. These clumps are typically at large distances and unresolved, so their physical properties cannot be reliably derived from the observations alone. Numerical simulations are needed to interpret the observations. To this end, we generate synthetic Herschel observations using our large-scale star-formation simulation, where massive stars explode as supernovae driving the interstellar-medium turbulence. From the synthetic observations, we compile a catalog of compact sources following the exact same procedure as for the Hi-GAL compact source catalog. We show that the sources from the simulation have observational properties with statistical distributions consistent with the observations. By relating the compact sources from the synthetic observations to their three-dimensional counterparts in the simulation, we find that the synthetic observations overestimate the clump masses by about an order of magnitude on average due to line-of-sight projection, and projection effects are likely to be even worse for Hi-GAL Inner Galaxy sources. We also find that a large fraction of sources classified as protostellar are likely to be starless, and propose a new method to partially discriminate between true and false protostellar sources.  

\end{abstract}


\section{Introduction}

Massive stars are essential constituents of the ecosystem of galaxies, driving the thermodynamical and chemical evolution of their interstellar medium (ISM). Understanding their formation process is a prerequisite for modeling the evolution of galaxies and investigating the high-redshift universe. A complete theory of star formation is not available yet, massive-star formation being perhaps the main hurdle to overcome. An important limitation in the study of massive stars is the difficulty to test our theoretical models against observational data, because massive stars are more rare and shorter lived than low-mass stars. Regions of massive star formation tend to be at relatively large distances, obscured by high extinction levels, and confused by complex gas and stellar dynamics, as massive stars form in stellar clusters. Observations of such regions can be properly interpreted only through synthetic observations of realistic theoretical models.  

Because of the stochastic nature of the dynamics of the ISM, synthetic and real observations of star-forming regions can only be compared statistically, which requires large samples. The ``Herschel Infrared Galactic Plane Survey'' (Hi-GAL) has produced the largest catalog to date of massive clumps \citep{Molinari+2016,Elia+2017,Elia+21}, usually viewed as potential progenitors of massive stars. Follow-up studies have characterized the dynamics of some of those clumps, including their infall rates \citep[e.g.][]{Traficante+17,Traficante+18}, providing important clues to the origin of massive stars. In this work, we compile the first synthetic catalog of massive clumps that can be compared statistically to the Hi-GAL compact source catalog, based on a star-formation simulation of a 250~pc region of the ISM driven by supernova (SN) explosions. 

Using earlier evolutionary stages of the same simulation, we have previously shown that SNe alone can drive the observed turbulence in MCs \citep{Padoan+SN1+2016ApJ,Pan+Padoan+SN2+2016ApJ} and can explain both the formation and dispersion of MCs \citep{Lu+20SN}. The star formation rate per free-fall time in the clouds was also found to be consistent with the observations \citep{Padoan+17sfr}. The simulation was then used to study the formation of massive stars in \citet{Padoan+20massive}, where it was also shown that observations could grossly overestimate the mass of protostellar cores, depending on distance and angular resolution. That study adopted a theoretician's perspective, by considering only the progenitor cores of massive stars, and only at the special moment when they have just started to collapse (the end of their prestellar phase). 

In this work, we adopt an observer's perspective, by selecting compact sources from synthetic observations of individual simulation snapshots, following the same procedure as in Herschel's Hi-GAL compact source catalog \citep{Elia+2017,Elia+21}. This approach results in a very large catalog of 52,543 synthetic sources, including both prestellar and protostellar ones, to be compared with the 150,223 sources of the Hi-GAL catalog. Our goal is twofold: to validate our synthetic catalog through the comparison with real observations and, once validated, to use it for the interpretation of the observations. We show that the clumps from the simulation have observational properties with statistical distributions consistent with the observations. We then compare the compact sources from the synthetic observations to their three-dimensional counterparts in the simulation. We find that the clump masses from the observations are generally overestimated due to line-of-sight projection and that a significant fraction of clumps classified as protostellar are likely to be starless.  

The structure of the paper is as follows. In \S~\ref{simulation}, we briefly summarize the numerical simulation. Radiative transfer and synthetic observations are presented in \S~\ref{synthetic}, while the procedure to compile the clump catalog from the synthetic observations is described in \S~\ref{sample}. The observational properties of the synthetic clumps are presented in \S~\ref{observational}, where they are also compared with the corresponding properties of the clumps in the Hi-GAL catalog. The synthetic sources are then compared with their three-dimensional counterpart from the simulation in \S~\ref{intrinsic}. Various implications of our results are discussed in \S~\ref{discussion}, and the main conclusions are summarized in \S~\ref{conclusions}.

\section{Simulation} \label{simulation}

This work is based on the same large-scale MHD simulation of star formation used in \citet{Padoan+17sfr} to study the star-formation rate in molecular clouds, in \citet{Padoan+20massive} to study the formation of massive stars, and in \citet{Lu+20SN} to study the effect of SNe on the dispersion of MCs. The simulation has been continuously run, during the past two years, under a multi-year PRACE project, and will be run for another year, until it reaches approximately 100~Myr of evolution. It describes an ISM region of size $L_{\rm box}=250$ pc and total mass $M_{\rm box}=1.9\times 10^6$ $\rm M_{\odot}$, where the turbulence is driven by SNe alone. The 3D MHD equations are solved with the AMR code RAMSES \citep{Teyssier+2002A&A,Fromang+06,Teyssier07}, using periodic boundary conditions. We refer the reader to the papers cited above for details about the numerical setup. In the following, we briefly summarize only the main features relevant to this work.

The energy equation includes the pdV work, the thermal energy introduced to model SN explosions, a uniform photoelectric heating as in \citet{Wolfire+95}, with efficiency $\epsilon=0.05$ and the FUV radiation field of \citet{Habing68} with coefficient $G_0=0.6$ (the UV shielding in MCs is approximated by tapering off the photoelectric heating exponentially above a number density of 200 cm$^{-3}$), and a tabulated optically thin cooling function constructed from the compilation by \citet{Gnedin+Hollon12} that includes all relevant atomic transitions. Molecular cooling is not included, due to the computational cost of solving the radiative transfer. The thermal balance between molecular cooling and cosmic-ray heating in dense gas is emulated by setting a limit of 10~K as the lowest temperature of dense gas. However, to generate synthetic observations of the dust emission, the radiative transfer is computed postprocessing individual snapshots, including all stars with mass $>2 \, \rm M_{\odot}$ as point sources (see \S~\ref{synthetic}).     

The initial conditions of the simulation are zero velocity, uniform density, $n_{\rm H,0}=5$ cm$^{-3}$, uniform temperature, $T_0=10^4$ K, and uniform magnetic field, $B_0=4.6$ $\mu$G. During the first 45~Myr, self-gravity was not included and SN explosions were randomly distributed in space and time, at a rate of 6.25 SNe Myr$^{-1}$. The resolution was $dx=0.24$ pc, achieved with a $128^3$ root grid and three AMR levels. The minimum cell size was then decreased to $dx=0.03$ pc, using a root-grid of $512^3$ cells and four AMR levels, for an additional period of 10.5 Myr, still without self-gravity. At $t=55.5$ Myr, gravity is introduced and the minimum cell size is further reduced to $dx=0.0076$ pc by adding two more AMR levels. This resolution allows us to resolve the formation of individual massive stars, so the time and location of SNe are computed self-consistently from the evolution of the massive stars. 

Individual stars are modeled with accreting sink particles, created when the gas density is larger than $10^6$ cm$^{-3}$ and other conditions are satisfied \citep[see][for details of the sink particle model]{Haugboelle+2018}. A SN is created when a sink particle of mass larger than 7.5~$\rm M_{\odot}$ has an age equal to the corresponding stellar lifetime for that mass \citep{Schaller+92}. The sink particle is removed and the stellar mass, momentum, and $10^{51}$ erg of thermal energy are added to the grid with a Gaussian profile \citep[see][for further details]{Padoan+SN1+2016ApJ}. By the last simulation snapshot used in this work, corresponding to a time of $34.2$~Myr from the inclusion of self-gravity and star formation, $4,283$ stars with mass $> 2\, \rm M_{\odot}$ have been generated, of which $\sim 415$ have already exploded as SNe. The stellar mass distribution is consistent with Salpeter's IMF \citep{Salpeter55} above $\sim 8\,\rm M_{\odot}$, but is incomplete at lower masses (it starts to flatten at a few solar masses instead of at a fraction of a solar mass), as expected for the spatial resolution of the simulation.

\section{Synthetic Observations} \label{synthetic}

\begin{figure}
\centering
\includegraphics[width=0.48\textwidth]{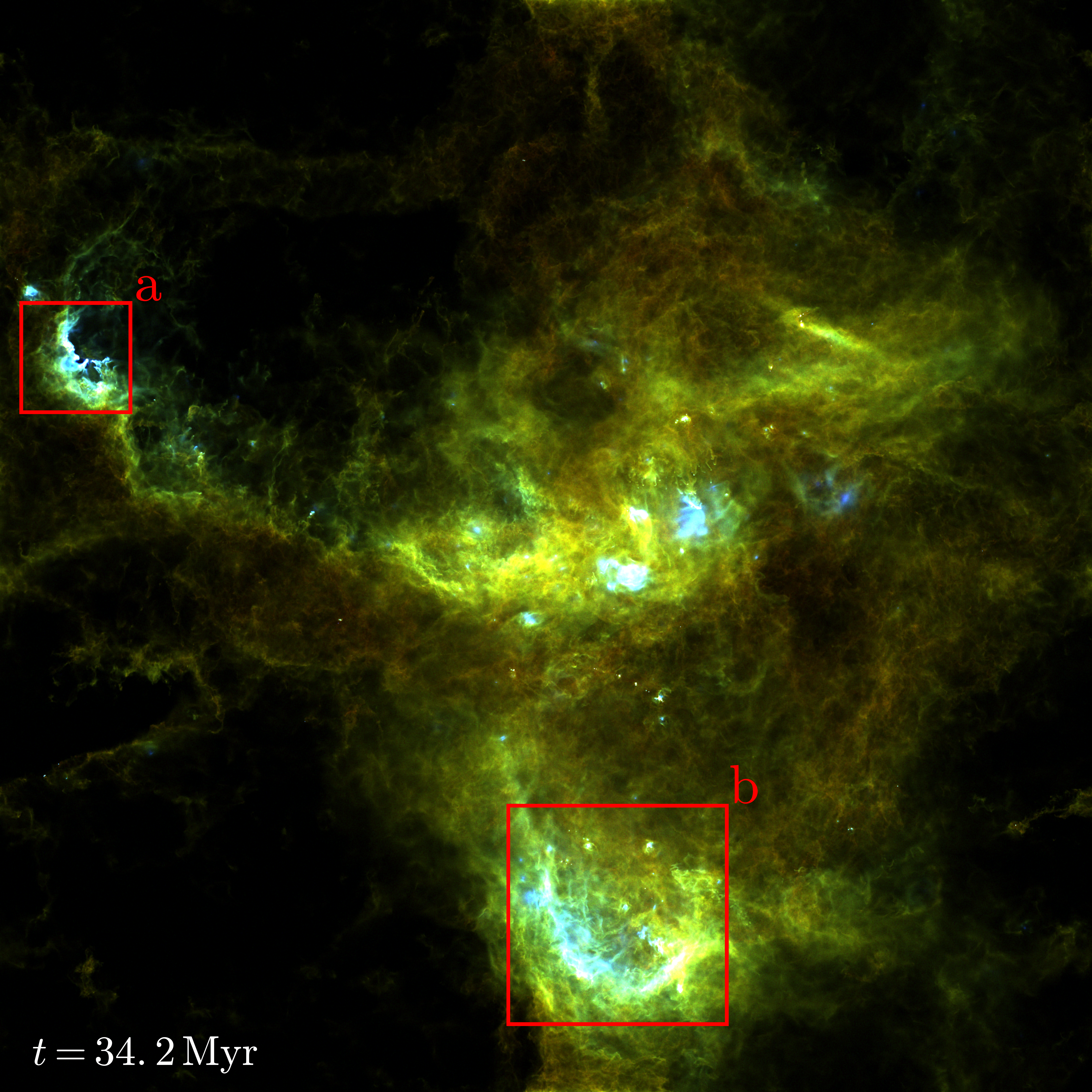}\\
\includegraphics[width=0.48\textwidth]{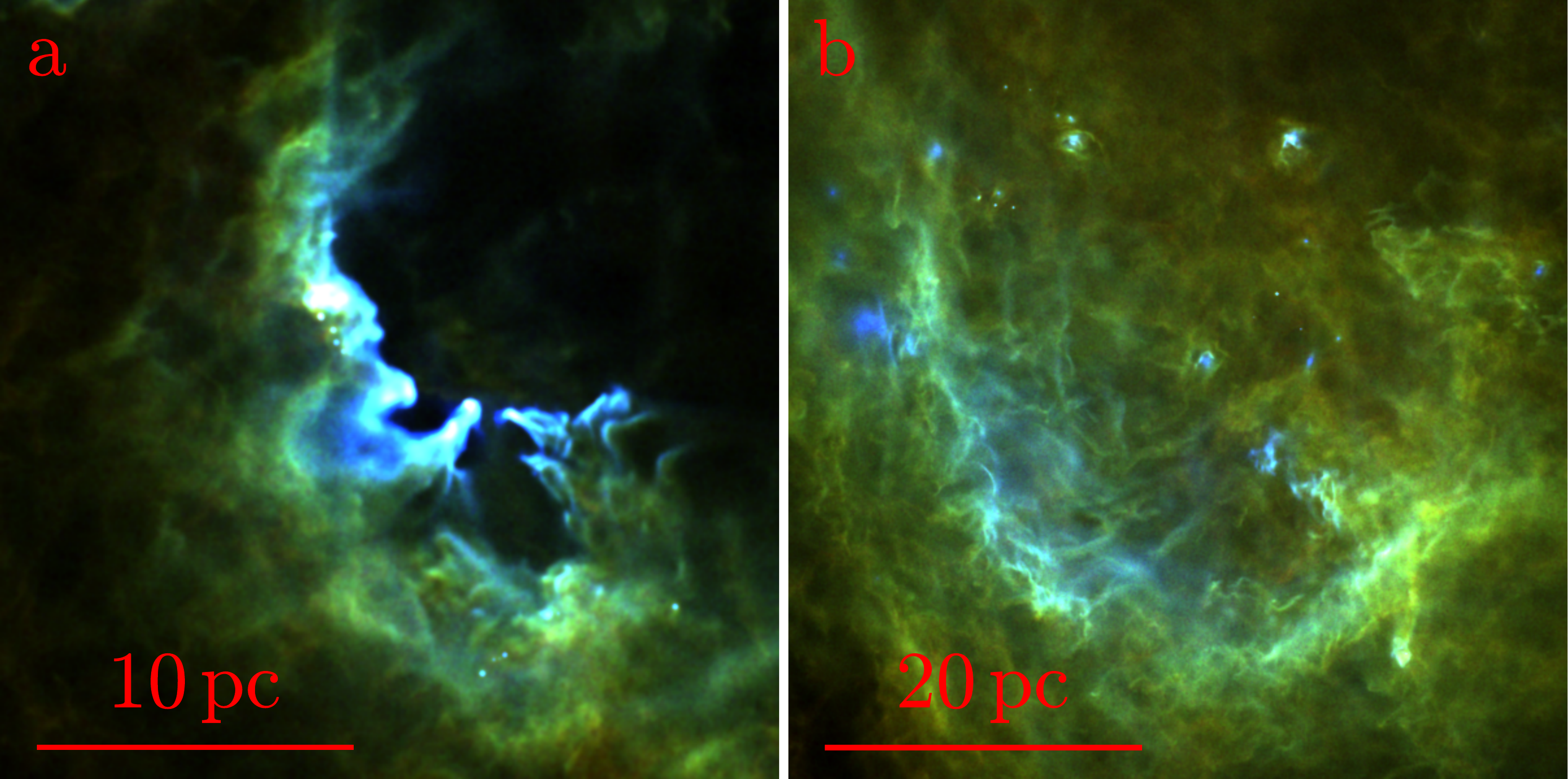}
\caption{{\it Upper panel}: Simulated Herschel's three-colour image of the whole 250 pc simulation at 34.2 Myr after the start of star formation, assuming a distance of 2~kpc, generated from single-band images at 70\,$\mu$m, 160\,$\mu$m, and 250\,$\mu$m for blue, green and red, respectively. The range of the linear colour scale is from 0 to 600~MJy\,sr$^{-1}$. The two red boxes mark the positions of the two zoom-in regions shown in the lower panels. {\it Lower panels}: The 25~pc (a) and 50~pc (b) zoom-in regions marked in the upper panel. The linear colour scale of these zoom-in regions is from 0 to 1200 MJy\,sr$^{-1}$.}
\label{fig_3color_map}
\end{figure}

To compute synthetic dust continuum maps, we select three snapshots of our simulation at times $15.4$, $23.3$, and $34.2 \,\rm Myr$ from the beginning of self-gravity and star formation. These three snapshots were chosen to sample different conditions in the star-formation history of the simulation. The first and the third times correspond to periods of relatively low star-formation rate, while the middle time corresponds to a peak in the star-formation rate. Furthermore, the first snapshot is early on in the simulation, where the star-formation efficiency is still relatively low, while the last snapshot is close to the end of the run, when the star-formation efficiency has increased significantly. We then calculate synthetic observations in Herschel's bands (70, 160, 250, 350, and 500 $\mu$m) to compare our results with Herschel's Hi-GAL compact-source catalog \citep{Elia+2017}.

The surface brightness maps were computed with the continuum radiative transfer program SOC \citep{Juvela_2019}.  The spatial discretization uses the full density information from the MHD run, with a root grid of 512$^3$ cells and six levels of refinement in the octree hierarchy. For the dust properties we tested both the diffuse-medium dust model of \citep{Compiegne_2011} and the dust model of \citep{OH_1994} that is more appropriate for dense medium. The final calculations were all carried out using the latter, which corresponds to dust grains with thin ice mantles, after $10^5$ years of coagulation at a density of 10$^6$\,cm$^{-3}$. This model yields a value of $\beta\approx 1.8$ for the exponent of the power-law dust emissivity. For the radiation that enters the model from the outside, we used the values for the normal local interstellar radiation field \citep{Mathis_1983}. 

As internal sources for the radiative transfer calculations, all stars with mass $> 2\, \rm M_{\odot}$ that have not exploded as SNe yet, were included as point sources, with their luminosities derived from the Zero Age Main Sequence mass-luminosity relations \citep{Duric+2004book, Salaris+2005book}, and their black-body spectra from calculating the effective temperature using the above luminosity and the mass-radius relations in \citet{Kippenhahn+1994book}. The number of stellar sources were 909, 2431, and 3868 for the three snapshots at times $15.4$, $23.3$, and $34.2 \,\rm Myr$ respectively.

SOC was used to calculate the equilibrium dust temperature for each model cell and, based on that information, the surface brightness maps at the Herschel frequencies. 
The surface brightness maps are resampled to the same pixel sizes as the Hi-GAL maps: 3.2, 4.5, 6.0, 8.0, and 11.5 arcsec, for the five bands in the order of increasing wavelength. At each wavelength, the full width at half maximum (FWHM) values of the adopted Gaussian telescope beams are three pixels, giving 9.6, 13.5, 18.0, 24.0, and 34.5 arcsec.
We added observational noise to the maps so that, after beam convolution appropriate for the assumed distances (2, 4, 8, and 12\,kpc), the noise was consistent with actual observations. For the surface brightness relative noise values, we assumed 4\% in the PACS bands (70\,$\mu$m and 160\,$\mu$m) and 2\% in the SPIRE bands (250, 350, and 500\,$\mu$m). 
We also added additional noise with absolute levels of 7.8, 6.0, 0.81, 0.42, and 0.28\,MJy\,sr$^{-1}$, to the five bands in the growing wavelength order. This noise was estimated by extracting small 100 arcsec by 100 arcsec submaps along overlap regions of two individual Hi-GAL tiles at Galactic longitudes l = $49^{\circ}$, $89^{\circ}$, and $144^{\circ}$. The paired submaps of the same region were used to calculate the rms of the surface brightness difference of each pixel. The rms noise of the low surface brightness submaps was consistent regardless of the region and was adopted as the absolute noise level in the synthetic observations. SOC is based on the Monte Carlo method, which also contributes to the noise in the maps. However, the number of simulated photon packages was chosen to be large enough so that the Monte Carlo noise is a few times below the observational noise.

For each of the three snapshots, the surface brightness maps were computed in the three major-axis directions, and assuming four different distances of 2, 4, 8, and 12\,kpc. This resulted in 36 maps at each wavelength: three snapshots seen from three orthogonal directions and at four assumed distances each. One of these 36 maps is shown in Figure~\ref{fig_3color_map} as a three-colour image of the whole 250 pc volume. The image is made by using the 70\,$\mu$m, 160\,$\mu$m, and 250\,$\mu$m maps for blue, green and red colors, respectively. The lower panels of Figure~\ref{fig_3color_map} show three-color images of two dense regions of 25 and 50~pc size, hosting the formation of massive stars.

\section{Synthetic Clump Catalogue} \label{sample}

Using our synthetic Herschel observations, we compile a synthetic catalogue of compact sources extracted with the CuTEx code \citep{Molinari+11}, with the same exact method used to generate the Hi-GAL compact source catalog \citep{Elia+2017,Elia+21}. All the steps of the source extraction are described in this section. The resulting synthetic clump catalog is summarized in Table~\ref{table_clump_properties} and described in Appendix~\ref{appA}.

\subsection{CuTEx}

\begin{figure}
\centering
\includegraphics[width=0.48\textwidth]{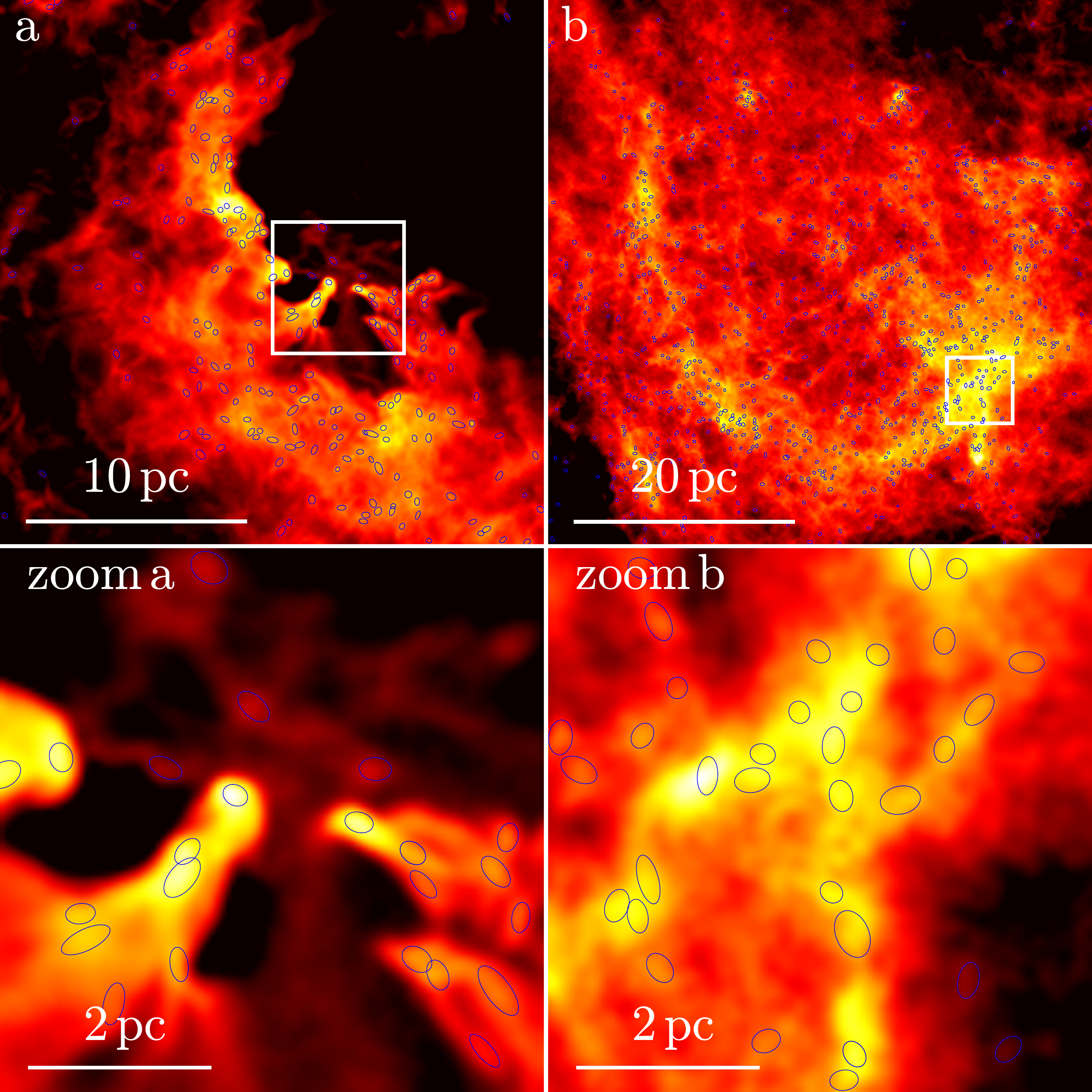}
\caption{{\it Upper panels:} 250~$\mu$m surface brightness of the two zoom-in regions from Fig.~\ref{fig_3color_map} (a: 25\,pc and b: 50\,pc). The colour scale is logarithmic with minimal values set equal to 1/32 (a) and 1/8 (b) of the maximum surface brightness in each region. The blue ellipses are the clumps detected by CuTEx at 250~$\mu$m, before band-merging. {\it Lower panels:} Additional 6~pc zoom-in regions corresponding to the white boxes in the upper panels. The logarithmic colour scale has the minimum values set at 1/32 (zoom a) and 1/3 (zoom b) of the maximum surface brightness in each zoom-in.}
\label{fig_singleband_image}
\end{figure}

CuTEx (CUrvature Thresholding EXtractor) is a detection and extraction code originally written in IDL and now also available in GDL. It is presented in detail in \citet{Molinari+11}, and summarized briefly here. CuTEx finds compact objects by calculating second order derivative maps, $\partial^2$, in four directions: along each axis and along the diagonals. The pixels that exceed a ``curvature" threshold value, $\zeta_{th}$, in all four maps are masked as possible sources, which are then clumped together into contiguous clusters. \citet{Molinari+11} report that using $\zeta_{th} \geq 0.5 \sigma_{\partial^2}$, where $\sigma_{\partial^2}$ is the rms in each particular $\partial^2$ map, the minimum number of contiguous pixels for reliable source detection is 3. The tentative location of the source is determined by finding a local maximum pixel that is $1 \sigma_{\partial^2}$ above the neighboring pixels in a map that is averaged from all four $\partial^2$ maps. If more local maxima pixels are found within the cluster, these become tentative other sources. If no statistically significant local maxima are found, the location is calculated as a mean of all the pixel coordinates in the cluster.

Once the source is located, the shape of the source is assumed to be a Gaussian. The size and the orientation is estimated from the local minima of the curvature around the source, using each of the four $\partial^2$ maps and measured along the gradient direction to find the minima before and after the source location, resulting in 8 measurements of the minima. If the before and after minima in each direction agree within 20\%, both are kept; otherwise only the minimum nearest to the source location is kept. An ellipse is fitted to the minima, and semi-major axis, semi-minor axis and position angle are recorded. Finally, if the semi-major axis is larger than three times the point spread function (PSF) of the observations, the size estimate is flagged as uncertain and set back to one PSF. This limits the size range of the initial guess from one PSF to three PSFs.

The source photometry is estimated by fitting a two-dimensional Gaussian shape of variable intensity, size and orientation, based on the initial guess on size and orientation. The background is estimated with a planar surface of a variable inclination and inclination direction, and is fit simultaneously with the source Gaussian, in a fitting area that is typically four times the point spread function (PSF) of the observations. If the source is in a crowded field, it and its nearest neighbors (typically within twice the PSF) are fitted simultaneously, although only the central source parameters are saved. The other, neighboring sources are fitted in their turn, based on their local background and neighbors. In this case of a crowded field, the fitting area is the minimum area that covers all the sources with an excess of one PSF around them. The Gaussian fitting routine is MPFIT \citep{Markwardt09}, which allows for the simultaneous adjusting of all source positions in a group, as well as varying the initial guesses on the source sizes by up to 30\%. If the initial guess was uncertain, the size is constrained only by the limits imposed by the photometric routine, that allows it to vary from a minimum of 0.95 PSF to the maximum of 3.9 PSFs, which can be achieved if the initial guess was already three PSFs. The photometric ASCII output file includes the source position, size, orientation, the integrated, peak and background fluxes, as well as uncertainties for all the photometric parameters. In addition, CuTEx creates a SAOImage region file of the detected source ellipses, which can be easily overplotted on an image, with examples shown in Figure~\ref{fig_singleband_image}. 

In this study, we use the same parameters for CuTEx as \citet{Molinari+2016}, with an extraction threshold  $\zeta_{th} = 2 \sigma_{\partial^2}$. Our synthetic maps are already resampled to the pixel resolution of three pixels per PSF, varying with the wavelength, similarly to the actual Herschel maps used by \citet{Molinari+2016}. We then perform a CuTEx detection at each of the five wavelength for the 3 selected snapshots seen from three orthogonal directions, and at 2, 4, 8 and 12 kpc distances. This results in 180 detection catalogues, which are used as inputs for the CuTEx photometric extraction routine to derive the photometry for the detected sources. In the end of the CuTEx detection and photometry, we have 180 single-band source catalogs for 36 different combinations of snapshots, viewing directions and distances.

\subsection{Band-merging and Final Source Selection} \label{merging}

As is in \citet{Elia+2017}, we band-merge the single-band catalogs, obtaining 36 multi-wavelength catalogues, one for each unique combination of snapshot, direction, and distance, using the following procedure:
\begin{itemize}
\item[(1)] Starting from the 500~$\mu$m sources, we seek detections at 350~$\mu$m that are a positional match by taking the center of the 350~$\mu$m sources, and checking which ones fall inside the circularized size of the 500~$\mu$m sources.
\item[(2)] We repeat the search for each lower wavelength by searching which 250~$\mu$m sources fall inside the circularized size of the 350~$\mu$m sources, which 160~$\mu$m sources fall inside the 250~$\mu$m sources, and finally which 70~$\mu$m sources are found inside the 160~$\mu$m sources.
\item[(3)] We select and retain only the band-merged sources that are detected in at least three consecutive Herschel bands (except for 70~$\mu$m band), meaning at 160–250–350~$\mu$m, at 250–350–500~$\mu$m, or at 160–250–350–500~$\mu$m, and without a dip in the SED between adjacent wavelengths or a peak at 500~$\mu$m \citep{Giannini+2012}.
\end{itemize}

\begin{figure*}
\centering
\includegraphics[width=1.\textwidth]{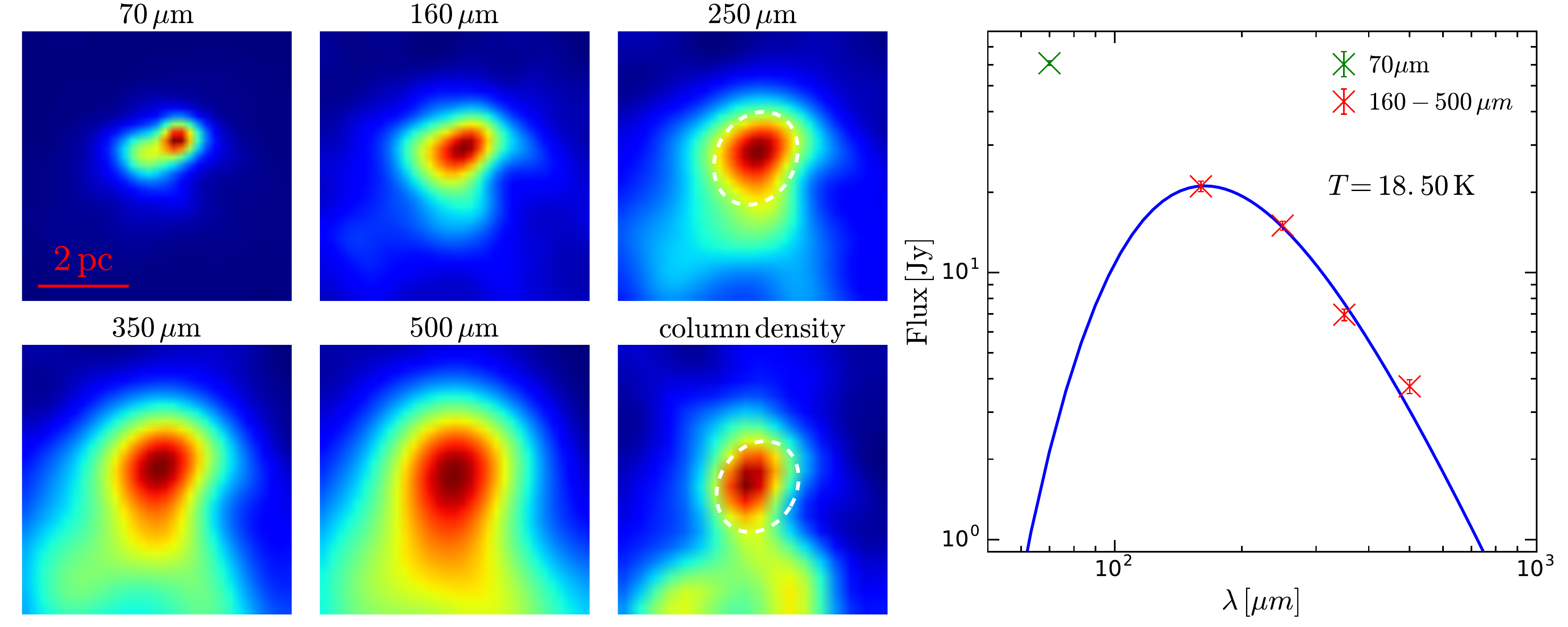}\\
\includegraphics[width=1.\textwidth]{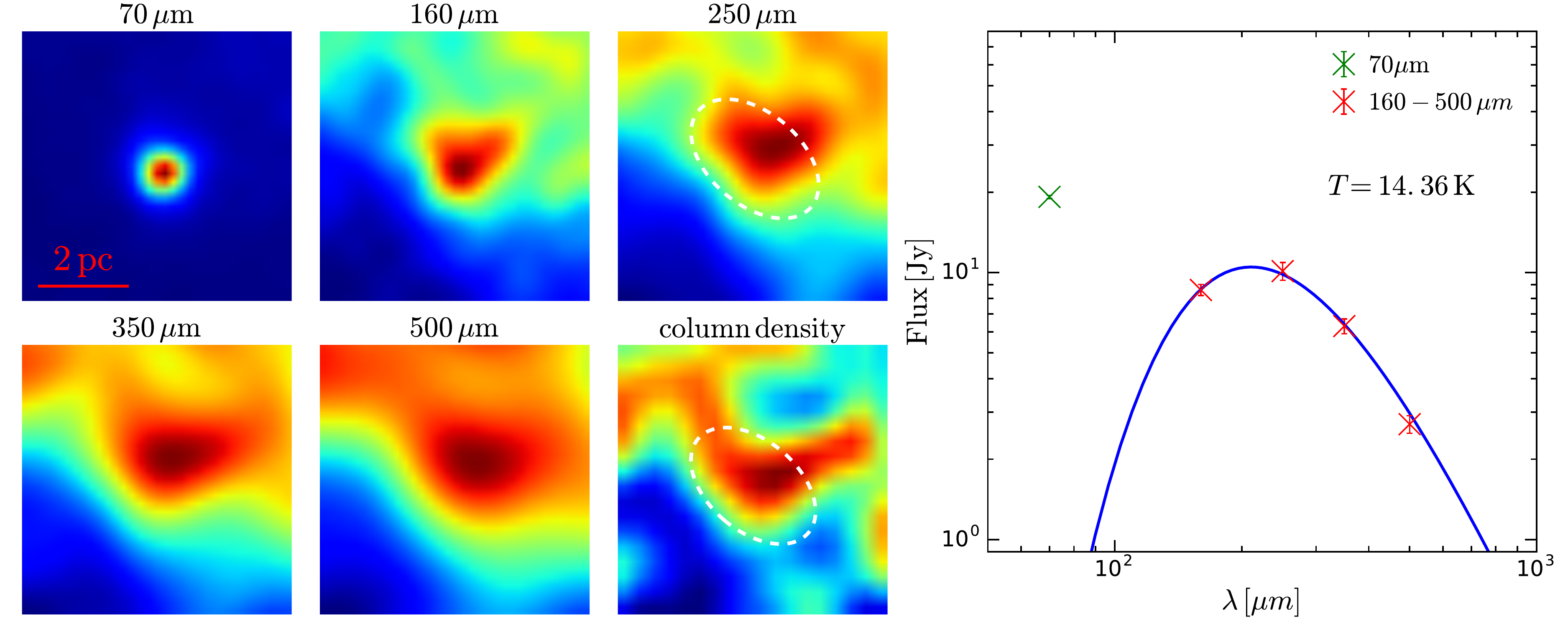}
\caption{Surface brightness at all five Herschel's bands and column density, in maps of 6~pc size, of two example sources assumed to be at a distance of 12~kpc. The dashed white ellipse is the CuTEx source detected at 250~$\mu$m. The corresponding SEDs are shown in the right panels, where the blue lines are the best fits. The 70~$\mu$m flux (green cross symbol), detected in both sources, is not included in the SED fitting.}
\label{fig_5band_image}
\end{figure*}

The circularized diameter of a source at each waveband is calculated as FWHM$_{\rm circ,\lambda}=\sqrt{\rm FWHM_{maj,\lambda}\times FWHM_{min,\lambda}}$, where $\rm FWHM_{maj,\lambda}$ and $\rm FWHM_{min,\lambda}$ are the semi-major and the semi-minor axis respectively of the ellipse of the source estimated by CuTEx. The beam-deconvolved diameter at each wavelength is estimated as $\theta_{\lambda}=\sqrt{\rm FWHM_{\rm circ,\lambda}^2-HPBM_{\lambda}^2}$, where $\rm HPBM_{\lambda}$ is the beam size at the given wavelength. However, if FWHM$_{\rm circ,\lambda} \leq {\rm HPBM}_{\lambda}$, we do not deconvolve. The fluxes at the wavelength $\lambda=$ 350 and 500~$\mu$m are then scaled by the ratio between the deconvolved sizes at those wavelengths and at 250~$\mu$m,
\begin{equation}
\overline{F}_{\lambda} =F_{\lambda}\times \frac{\theta_{250}}{\theta_{\lambda}}, \,\,\,\,\,\lambda\geq 350~\mu m,
\label{eq_flux_scaling}
\end{equation}
if $\theta_{\lambda}>\theta_{250}$, FWHM$_{\rm circ,\lambda}>{\rm HPBM}_{\lambda}$, and FWHM$_{\rm circ,250}>\rm HPBM_{250}$ \citep{Elia+2013,Elia+2017}. If these conditions are not fulfilled, then we do not perform any flux scaling at that wavelength. The 250~$\mu$m wavelength is taken as the reference wavelength because it is the shortest one that is always present when it is imposed that the sources have detections for at least three consecutive wavelengths. Using that reference wavelength, the source diameter, $D$, in the catalog is either the beam-deconvolved diameter, $D=\theta_{250}$, if FWHM$_{\rm circ,250}>{\rm HPBM}_{250}$, or the circularized diameter, $D={\rm FWHM}_{\rm circ,250}$, if FWHM$_{\rm circ,250}<{\rm HPBM}_{250}$. The error of the fluxes are estimated by the product of the area of the circularized clump size and the the RMS of the residual pixel fluxes after the subtraction of the best fit to the initial data by CuTEx.

\subsection{SED Fitting}\label{SED_fitting}

As in \citet{Elia+2017}, we use a single greybody function to fit the $\lambda$ $\geq$ 160~$\mu$m SEDs from the CuTEx photometric output to estimate the temperature of the sources, $T$, and thus derive their masses, $M_{\rm SED}$, 

\begin{equation}
F_{\nu}=(1-\mathrm{e}^{-\tau_{\nu}})\,B_{\nu}(T)\,\Omega\,,
\label{eq_greybody}
\end{equation}
where $F_{\nu}$ is the observed flux density at the frequency $\nu$, $\tau_{\nu}$ is optical depth of the medium, $B_{\nu}(T)$ is the Planck's function at the dust temperature $T$, and $\Omega$ is the source solid angle in the sky. The optical depth is parametrised as:
\begin{equation}
\tau_{\nu}=\left( \frac{\nu}{\nu_0} \right)^{\beta} \,,
\label{eq_tau}
\end{equation}
where the parameter $\nu_0=c/\lambda_0$ is such that $\tau_{\nu_0}=1$ and $\beta$ is the exponent of the power-law dust emissivity at large wavelengths. While \citet{Elia+2017} adopt the value $\beta=2$, we use $\beta=1.8$, which is more appropriate for the dust model of \cite{OH_1994} that we have used in the radiative transfer. By keeping self-consistency between the dust model and the SED fitting, we avoid the introduction of artificial discrepancies between the mass in the model and that deduced from the SED fitting. $\Omega$ is taken to be equal to the surface area measured by CuTEx at 250~$\mu$m, and we find the values of $T$ and $\lambda_0$ from a least-square-fit of the SED using Equation~\ref{eq_greybody}, within the ranges 5~K$\le T\le 40$~K and 5~$\mu$m$\le \lambda_0 \le 350\mu$m, as in \cite{Elia+2017}. For a small number of sources the best-fit temperature would be larger than 40~K, but it is forced to be equal to 40~K as that is the maximum value of the temperature range adopted for the fit. To avoid uncertainties related to these sources, we drop all sources with $T=40$~K from both the Hi-GAL catalog (396 sources) and the synthetic catalog (311 sources).   

The mass is then derived as
\begin{equation}
M_{\rm SED} = (d^2 \Omega/\kappa_{\rm ref})\tau_{\rm ref}\,,
\label{eq_mass}
\end{equation}
where $\kappa_{\mathrm{ref}}$ is the opacity estimated at a reference frequency $\nu_{\mathrm{ref}}$. 
As in \citet{Elia+2017}, we adopt $\lambda_{\rm ref}=300$~$\mu$m and $\kappa_{\mathrm{ref}}=0.1$~cm$^2$~g$^{-1}$. 
This value of the opacity is only $\approx 30$\% lower than in the dust model of \cite{OH_1994}.  

If $\lambda_0 \le 44.5$~$\mu$m, then $\tau\le 0.1$ at 160~$\mu$m, and the SEDs are instead fitted with the optically thin expression for the flux density,
\begin{equation}
F_{\nu} = \frac{M_{\rm SED}\,\kappa_{\mathrm{ref}}}{d^2}\,\left(\frac{\nu}{\nu_{\mathrm{ref}}}\right)^\beta\,B_{\nu}(T)\,,
\label{eq_greybody_thin}
\end{equation}
which gives directly the values of both $T$ and $M_{\rm SED}$.

\subsection{Source Classification}\label{classification}

Following \citet{Elia+2017}, we classify the synthetic compact sources in our catalog as protostellar if they have a CuTEx counterpart at 70~$\mu$m, or starless if they do not. 
Single waveband images, column density maps (at the resolution of the 250~$\mu$m observations) and the corresponding SEDs are shown in Figure~\ref{fig_5band_image} for two example sources at 12~kpc distance. The 70~$\mu$m flux is not used in the SED fitting procedure, but only to classify the sources as protostellar or starless. As shown by the right panels of Figure~\ref{fig_5band_image}, both sources have a strong 70~$\mu$m excess, so they are classified as protostellar. As discussed later in \S~\ref{real_mass} and \S~\ref{protostellar}, the source in the upper panels is an example of a real protostellar clump, while the one in the lower panels is primarily a projection effect and a ``false" protostellar source.

\begin{table}
\centering
\caption{Number of sources, divided into starless and protostellar categories, and median values of diameter, mass, and temperature at different distances. (The total number of sources is 52,232, but becomes 108,102 when we normalize the numbers for distances $> 2$~kpc to match the distance distribution in the Hi-GAL Inner Galaxy catalog as described in \S~\ref{sect_DMT}).}
\begin{tabular}{lcccc}
\hline
\hline
& $N$ & $\rm Diameter $ & $\rm Mass $ & $\rm Temperature $ \\
& & $\rm [pc]$ & $\rm [M_{\odot}]$ & $\rm [K]$ \\
\hline
\,\,2 kpc  &  36422  &  0.23  &  11.81  &  11.88 \\
\,\,\,\,\,\,\,starless  &  32146  &  0.23  &  12.41  &  11.52 \\
\,\,\,\,\,\,\,protostellar  &  4276  &  0.23  &  7.28  &  16.84 \\
\hline
\,\,4 kpc  &  11415  &  0.45  &  41.10  &  12.69 \\
\,\,\,\,\,\,\,starless  &  9285  &  0.46  &  44.91  &  12.17 \\
\,\,\,\,\,\,\,protostellar  &  2130  &  0.43  &  25.75  &  17.18 \\
\hline
\,\,8 kpc  &  3026  &  0.89  &  152.62  &  13.60 \\
\,\,\,\,\,\,\,starless  &  2111  &  0.90  &  180.23  &  12.59 \\
\,\,\,\,\,\,\,protostellar  &  915  &  0.84  &  104.44  &  17.41 \\
\hline
12 kpc  &  1369  &  1.32  &  338.66  &  14.14 \\
\,\,\,\,\,\,\,starless  &  868  &  1.38  &  405.96  &  12.83 \\
\,\,\,\,\,\,\,protostellar  &  501  &  1.24  &  241.28  &  17.61 \\
\hline
all  &  52232  &  0.26  &  16.74  &  12.21 \\
\,\,\,\,\,\,\,starless  &  44410  &  0.26  &  17.10  &  11.75 \\
\,\,\,\,\,\,\,protostellar  &  7822  &  0.30  &  15.26  &  17.04 \\
\hline
\hline
\end{tabular}
\label{table_clump_properties}
\end{table}

\section{Observational Properties of Synthetic Clumps} \label{observational}

In this section, we compute the observational properties of the compact sources from our synthetic observations and compare them with those of the sources in Hershel's Hi-GAL catalog \citep{Elia+21}.
Our main purpose is to validate our synthetic clump catalog, before using it to interpret the observations in the following section. The number of synthetic sources, and the median values of their diameters, masses and temperatures at the different distances are summarized in Table~\ref{table_clump_properties}.

Although the star-formation rate in the simulation is consistent with both the Kennicutt-Schmidt relation, globally, and with the observed star-formation rate in molecular clouds, at smaller scales \citep{Padoan+12sfr}, its size, 250~pc, and mean column density, 30~$\rm M_{\odot}$pc$^{-2}$, makes the synthetic observations more suitable for comparison with a single spiral arm than with the entire Galactic plane (e.g. the column density of the Perseus arm is estimated to be $\sim 23 \,\rm M_{\odot}\,{\rm pc}^{-2}$ \citep{Heyer+Terebey98}). Thus, for the purpose of validating the synthetic catalog, we first compare it with the Hi-GAL catalog of the Outer Galaxy, defined by the Galactic longitude range $67^{\circ} - 289^{\circ}$ \footnote{\citet{Elia+21} adopted a galactocentric definition of the Inner/Outer Galaxy, separated by the Solar circle. However, due to our desire to avoid strong projection effects, we use the longitudinal definition of Inner/Outer Galaxy, as was done in \citet{Elia+2017}.}. In the case of the Outer Galaxy, a large fraction of sources are located in the outer Perseus arm, and projection effects should not be much stronger than in our simulation, as the number of sources drops almost exponentially with increasing distance. In \citet{Padoan+SN3+2016ApJ} it was already found, through synthetic CO observations, that molecular clouds extracted from the simulation are consistent with those from the FCRAO CO Survey of the Outer Galaxy \citep{Heyer+98}. 

Besides the comparison with the Outer Galaxy, we also carry out a parallel comparison with the Hi-GAL catalog of the Inner Galaxy ($0^{\circ} - 67^{\circ}$ and $289^{\circ} - 360^{\circ}$ longitude), in order to highlight potential differences arising from the much larger depth of the observations. The number of sources in the Inner Galaxy is nearly constant between $\sim 1.5$~kpc and $\sim 13.5$~kpc, so projection effects are expected to be significantly enhanced.        

In the comparison with the Hi-GAL catalogs, we distinguish between protostellar and starless sources. The fraction of protostellar sources, as a function of distance, is shown in the top panels of Figure~\ref{fig_distance} for both the synthetic catalog (solid line) and the Hi-GAL catalogs (dashed lines). In our catalog, the fraction increases slightly with increasing distance at all distances, while in the Hi-GAL catalogs it remains approximately constant, except for a slight increase in the Inner Galaxy catalog (top right panel) for distances larger than approximately 7~kpc. 

Two competing effects contribute to the distance dependence of the protostellar fraction. On the one hand, at larger distances neighboring sources may blend together. If a blended source consists of a starless and a protostellar source, it is marked as protostellar due to the detected 70~$\mu$m emission, with the effect of increasing the protostellar fraction with increasing distance. On the other hand, protostars of increasingly smaller mass and luminosity can be detected with decreasing distance, which tends to increase the protostellar fraction toward smaller distances and to cancel the effect of blending. While both effects are present in the observations, explaining the approximately constant protostellar fraction, the second one is not fully captured in the synthetic observations. Although we include as radiation sources all the stars down to 2~$\rm M_{\odot}$, the stellar IMF in our simulation is incomplete below $\sim$ 8~$\rm M_{\odot}$. Thus, we underestimate the relative number of low mass stars that may be potentially detected as low luminosity protostars at small distances, explaining the distance dependence of the protostellar fraction in the synthetic catalog. 

The protostellar fraction in the synthetic catalog is very close to that of the Outer Galaxy at intermediate distances (approximately 6 to 9~kpc), with a value of $\sim 0.3$. It is also the same as that of the Inner Galaxy at 6~kpc.

\subsection{Size, Mass and Temperature Distributions}
\label{sect_DMT}

\begin{figure}
\centering
\includegraphics[width=0.48\textwidth]{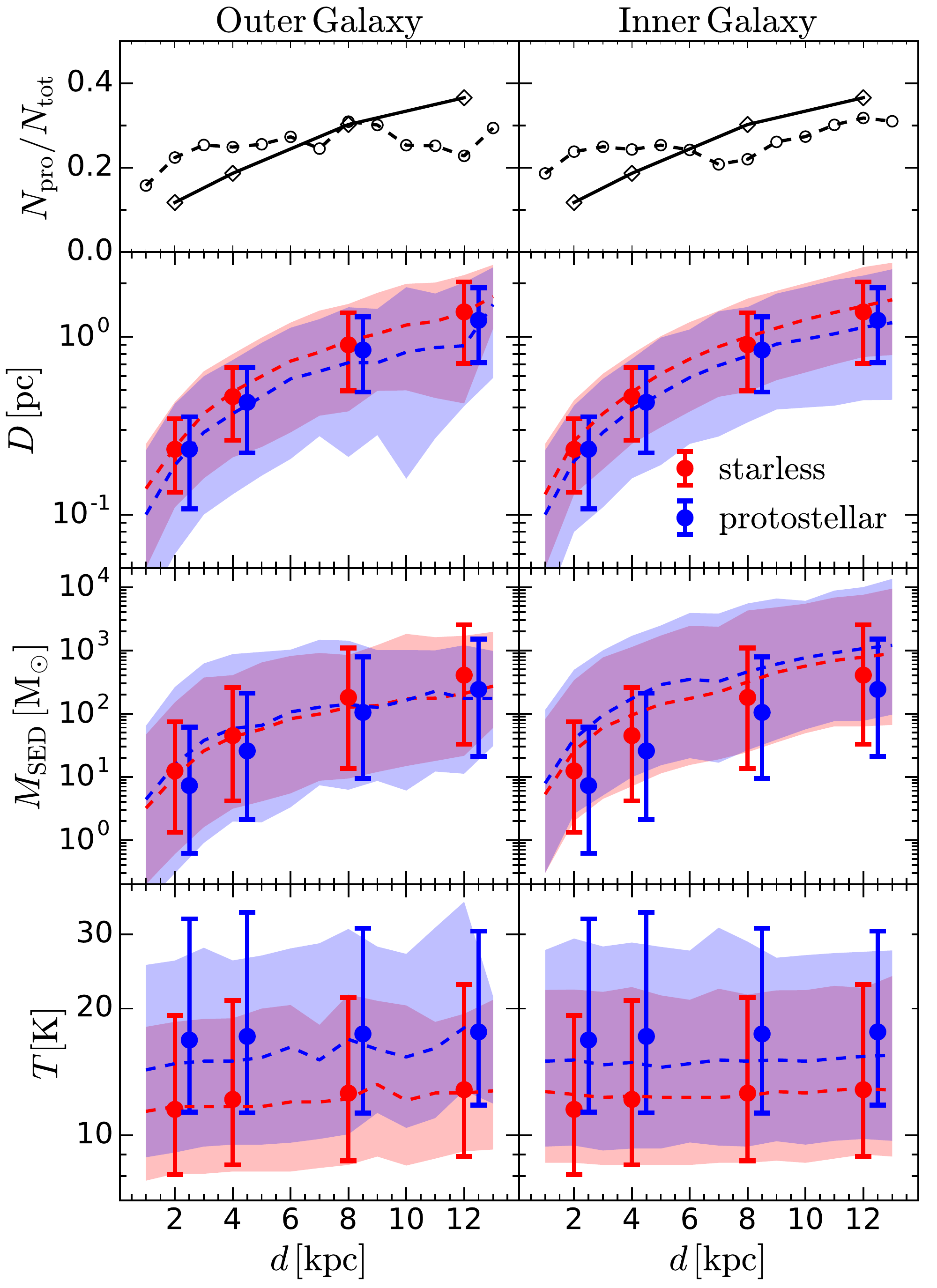}
\caption{Source classification and derived source parameters as a function of distance. {\it Top panels:} fraction of protostellar sources in the synthetic catalog (solid line) and in the Hi-GAL catalogs (dashed lines) for the Outer Galaxy (left) and the Inner Galaxy (right).  {\it Second row}: median values of the synthetic source diameters, separated into starless (red) and protostellar (blue) sources. Error bars are between the 2.5 and 97.5 percentiles. The median values for the Hi-GAL sources are shown by the dashed lines, while the shaded regions correspond to the same percentile range as for the error bars. {\it Third row}: the same as the second row of panels, but for the source mass derived from the SED fitting. {\it Fourth row}: the same as the previous row, but for the source temperature.}
\label{fig_distance}
\end{figure}

\begin{figure*}
\centering
\includegraphics[width=1.\textwidth]{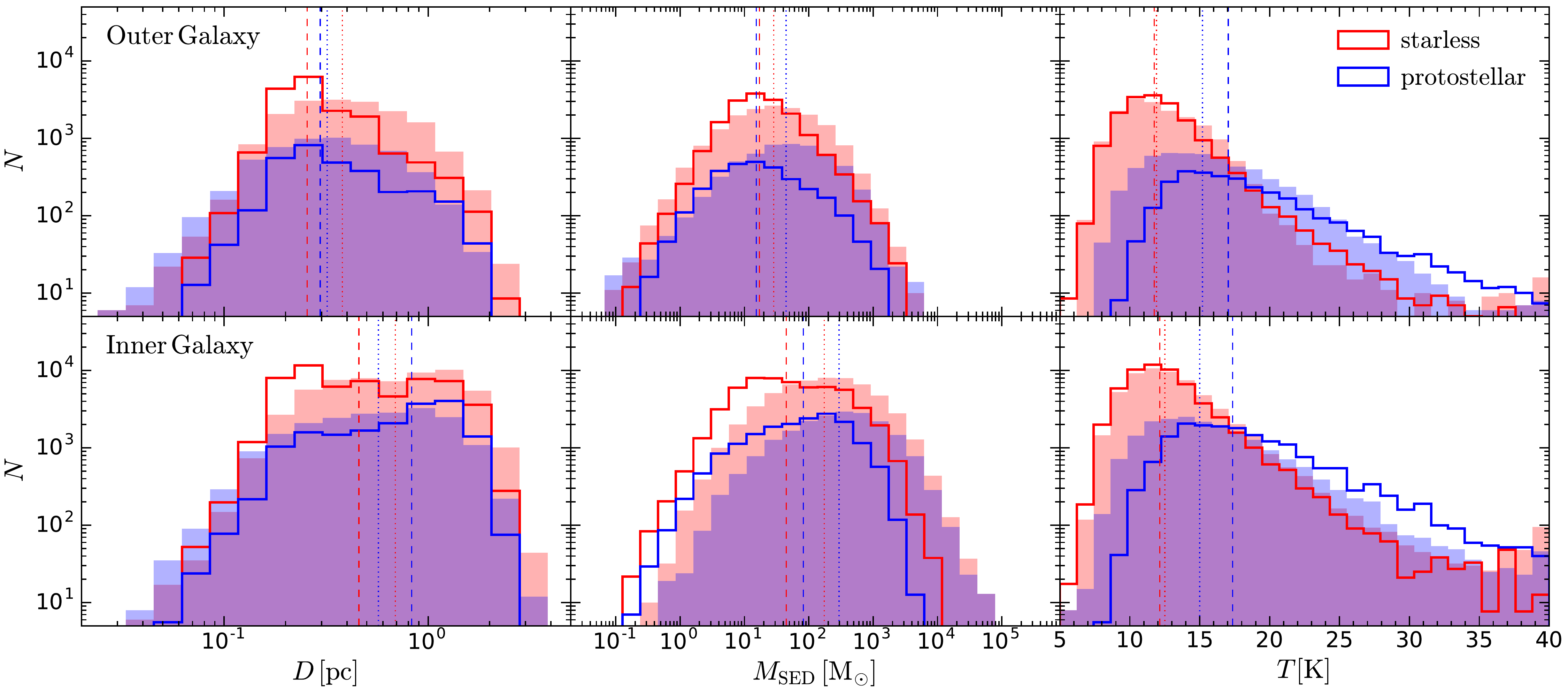}
\caption{{\it Upper panels}: distributions of the source diameters, masses, and temperatures for the starless (red solid line) and protostellar (blue solid line) sources from the synthetic catalog, compared to the sources from the Hi-GAL Outer Galaxy catalog (red and blue bars) in the distance interval between 1.5 and 13.5~kpc. The number of synthetic sources (both starless and protostellar) is normalized by the ratio of Hi-GAL and synthetic starless sources. The vertical dashed and dotted lines are the median values for the synthetic and Hi-GAL catalogs, respectively. {\it Lower panels}: the same as the upper panels, but comparing with the Hi-GAL Inner Galaxy catalog. To approximate the distance distribution of the Hi-GAL sources in the Inner Galaxy, the number of synthetic sources at distances $d>2$~kpc is scaled by a factor $(d/2 \, {\rm kpc})^2/2$.}
\label{fig_diameter}
\end{figure*}

The three bottom rows of panels of Figure~\ref{fig_distance} show the distance dependence of the source diameters, masses and temperatures as error-bar plots for the synthetic sources at distances 2, 4, 8 and 12~kpc, and as shaded areas for the Hi-GAL sources in bins of 1~kpc between 0.5~kpc and 13.5~kpc. The red and blue error bars give the median and the dispersion of the diameters, masses, or temperatures of our starless and protostellar sources respectively. The dispersion is estimated as the interval between the 2.5 and 97.5 percentiles. The red and blue dashed lines and shaded areas give the median and the dispersion for the starless and protostellar sources in the Hi-GAL catalogs. 

Figure~\ref{fig_distance} shows that our synthetic observations reproduce well the range of clump sizes found in Herschel's sources and their distance dependence, as well as the lower median values of the diameters of the protostellar clumps relative to that of the starless ones. The synthetic catalog reproduces well also the median values of the observed masses as a function of distance, and their dispersion, in the case of the Outer Galaxy, although it underestimates by a factor of 2-3 the masses of protostellar sources at 2 and 4~kpc. Compared with the Inner Galaxy, instead, our source masses, particularly the protostellar ones, are significantly lower than in the observations at all distances, a discrepancy that we attribute primarily to the much longer lines of sight through the Inner Galaxy (several kpc) than through our simulation (250~pc), or through the Outer Galaxy (dominated by the Perseus arm), as discussed later in \S~\ref{caveats}. We will argue that the difference between intrinsic and projected clump masses that will be discussed in \S~\ref{real_mass} must be even larger in the Inner Galaxy survey than in the synthetic observations. Finally, the bottom panels of Figure~\ref{fig_distance} show that the observed clump temperatures are well reproduced by the synthetic catalog, except for slightly larger values in the case of protostellar sources at 2 and 4~kpc in the Outer Galaxy, and at all distances in the Inner Galaxy. This temperature difference is discussed below in the context of Figure~\ref{fig_diameter}. The temperature of starless sources is instead perfectly reproduced in both median values and dispersion at all distances for both the Outer and the Inner Galaxy.  

Figure~\ref{fig_diameter} shows the probability distribution of the clump diameters (left panels), clump masses (middle panels), and clump temperatures (right panels) for starless and protostellar synthetic sources (solid-line histograms), compared with the same distributions for the Hi-GAL sources (shaded-area histograms) in the Outer Galaxy (upper panels) and in the Inner Galaxy (lower panels). All Hi-GAL sources with distances between 1.5 and 13.5\,kpc are included in the shaded-area histograms. Because the number of synthetic sources decreases with distance approximately at the same rate as in the Outer Galaxy catalog, the histograms from the synthetic catalog in the upper panels are computed using the actual number of sources extracted at each distance, and then normalizing the total number of synthetic starless sources to the same total number of observational starless sources, and using the same normalization factor for the synthetic protostellar clumps, too. In the case of the comparison with the Inner Galaxy (lower panels), the number of the synthetic sources is first multiplied by a factor $(d/2 \, {\rm kpc})^2/2$, for distances $d>2$~kpc (and left unchanged for sources at 2~kpc), to approximate the nearly constant number of sources in the Hi-GAL Inner Galaxy catalog at distances between 2 and 12~kpc. After this normalization by distance, the total number of synthetic sources has been normalized by the ratio of observed and synthetic starless sources. Thus, the area below the histograms of synthetic and observed starless sources are the same, while the areas below the histograms of protostellar sources reflect the actual ratios of protostellar to starless sources in both the synthetic and Hi-GAL catalogs. The vertical dashed and dotted lines are the median values of the synthetic and Hi-GAL catalogs, respectively. 

As shown by the left panels of Figure~\ref{fig_diameter}, there is very good agreement between the synthetic and Hi-GAL sources with respect to the source diameters in both the median values and the shapes of the probability distributions. The difference in the shape of the histograms between the Inner and Outer Galaxy is due to the different distance distributions of the sources. The more uniform distance distribution of the Inner-Galaxy sources relative to the Outer-Galaxy ones causes their flatter diameter distribution in the approximate diameter range between 0.2 and 2~pc, while the Outer-Galaxy sources are picked around 0.3~pc.

The central upper panel of Figure~\ref{fig_diameter} shows that the mass distribution of the synthetic starless sources fits well the corresponding distribution from the Outer Galaxy, while that of the protostellar sources slightly underestimates the number of massive sources. In the case of the Inner-Galaxy comparison (central lower panel of Figure~\ref{fig_diameter}), both the starless and the protostellar mass distributions of the synthetic sources are clearly shifted to smaller masses relative to the observations (see discussion in \S~\ref{caveats}).   

Finally, the right panels of Figure~\ref{fig_diameter} show that the temperature distributions of starless sources are nearly perfectly matched by the synthetic sources, both for the Outer and the Inner Galaxy. On the other hand, the temperatures of the synthetic protostellar sources have median values $\sim 2$~K larger than in the observations, as well as high-temperature tails that are a bit shallower than those of the Hi-GAL sources. These differences in the temperature distributions of protostellar sources are the expected consequence of the incompleteness of our stellar mass distribution. As mentioned in \S~\ref{simulation}, the stellar mass distribution in the simulation is consistent with Salpeter's IMF \citep{Salpeter55} above $\sim 8\,\rm M_{\odot}$, but is incomplete at lower masses (and no stars below $\sim 2\,\rm M_{\odot}$ are included as stellar sources in the radiative transfer calculation). At the lower distances, the 70~$\mu$m flux of most stars in the range $\sim 2-8\,\rm M_{\odot}$ would be detected in the observations, but many of these 70~$\mu$m sources would be missing in our simulation. Because the fraction of missing stars increases with decreasing masses, we miss preferentially colder sources, which matches the discrepancy in the temperature distributions of protostars in the right panels of  
Figure~\ref{fig_diameter}.

\subsection{Size, Mass and Temperature Correlations} \label{correlations}

\begin{figure*}
\centering
\includegraphics[width=0.48\textwidth]{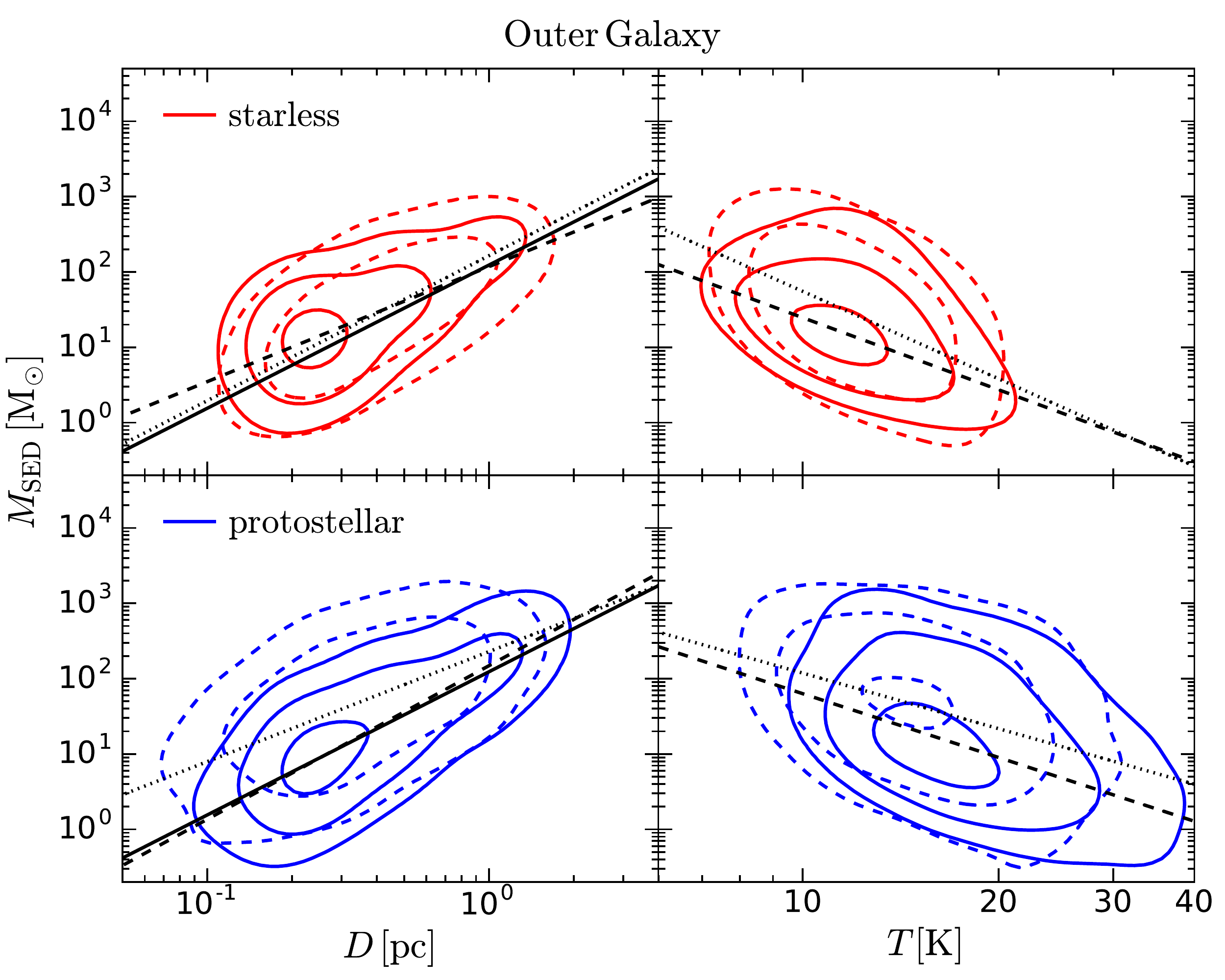}
\includegraphics[width=0.48\textwidth]{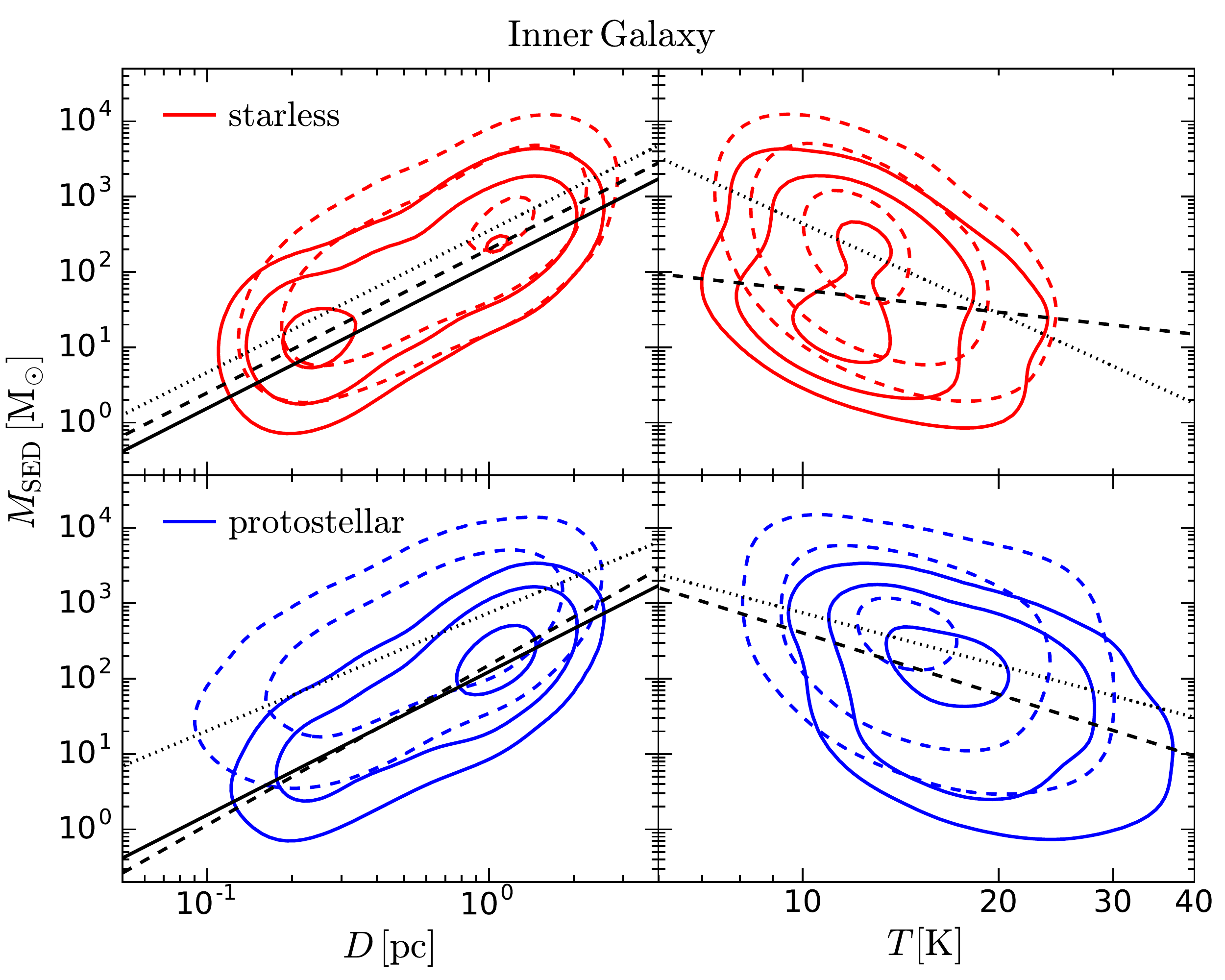}
\caption{{\it Left four panels}: mass versus diameter and mass versus temperature for starless (upper panels) and protostellar (lower panels) synthetic sources (solid contour lines). The black line is Larson's mass-size relation, $M(\rm r)=460~\rm M_{\odot}(r/pc)^{1.9}$ \citep{Larson81}. Source number density isocontours representing the starless and protostellar sources from the Hi-GAL Outer Galaxy catalog are also displayed with dashed lines. The number densities have been computed subdividing the area of the plot with a grid of $70 \times 70$ cells in logarithmic intervals. Once the global maximum of both distributions in the same panel (starless or protostellar) has been found, the plotted contours are chosen to show the number density of 5\%, 20\% and 70\% of the maximum, as in Figure~7 of \citet{Elia+2017}. {\it Right four panels}: the same as in left four panels, but comparing with the sources from the Hi-GAL Inner Galaxy survey. In this case the number of synthetic sources at distances $d>2$~kpc has been rescaled by a factor $(d/2 \, {\rm kpc})^2/2$, to mimic the distance distribution of the sources in the Inner Galaxy, as in Figure~\ref{fig_diameter}.}
\label{fig_correlations}
\end{figure*}

The left group of four panels of Figure~\ref{fig_correlations} shows the relation between the mass and diameter (left), and the mass and temperature (right) of the synthetic sources (solid contour lines) and of the Hi-GAL Outer Galaxy sources (dashed contour lines) for starless (upper panel) and protostellar (lower panel) sources separately. The right group of four panels shows the same, but for the Hi-GAL Inner Galaxy survey. The contours are lines of equal number density of sources. The number density is computed separately for the synthetic and Hi-GAL sources, by dividing the plane into $70 \times 70$ logarithmic intervals of diameter and mass. The contours correspond to 5\%, 20\% and 70\% of the maximum number density among all the cells from both samples. The comparison shows that our synthetic sources follow very closely the mass-diameter and the mass-temperature relations of the Outer Galaxy sources, with respect to both the contour lines and the median least-square-fit relations (dashed-dotted and dotted lines). 

In the case of the Inner Galaxy, as shown by previous figures, the observed clump masses are on average a few times higher than those of the synthetic observations, the discrepancy being stronger in the case of protostellar sources. In both Hi-GAL catalogs, the median mass-size relations have nearly the same slope as Larson's mas-size relation (solid line), corresponding to roughly constant surface density \citep{Larson81}. This slope is well reproduced in the synthetic catalog as well.       

All the mass-temperature contour plots and median relations from our catalog and from the observations show the same trend of decreasing mass with increasing temperature. However, this trend is not as significant as the correlation between mass and diameter, as shown by the values of Pearson's correlation coefficients, between $\sim -0.4$ and $\sim -0.5$ (compared to $\sim 0.7-0.8$ for the mass-size relations). It is possible that most of the anti-correlation between mass and temperature originates from the uncertainty in the temperature estimate from the SED fitting, rather than from a real physical anti-correlation between mass and temperature.

\section{Observational Versus Intrinsic Clump Properties} \label{intrinsic}

\begin{figure*}
\centering
\includegraphics[width=0.49\textwidth]{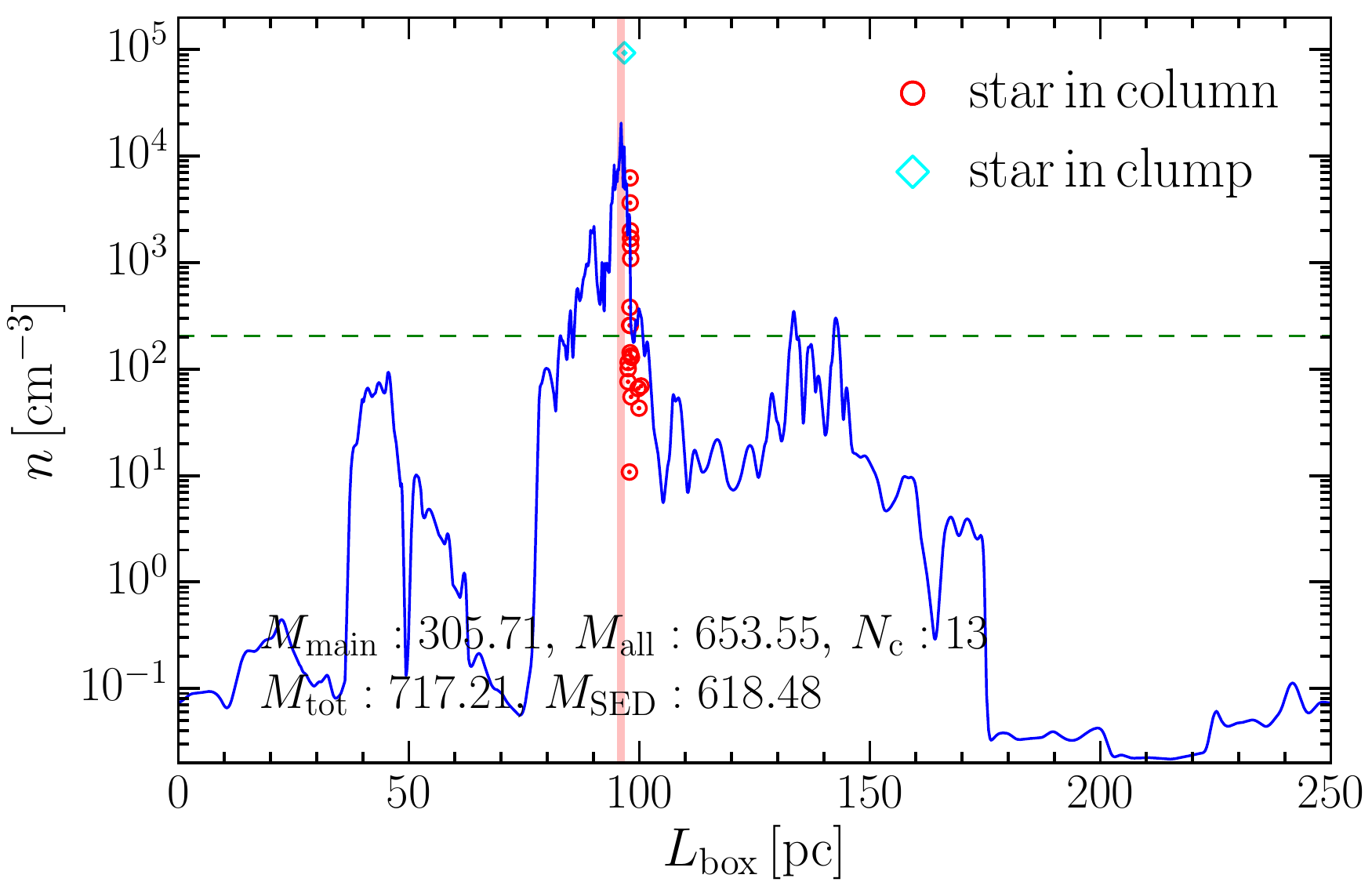}
\includegraphics[width=0.49\textwidth]{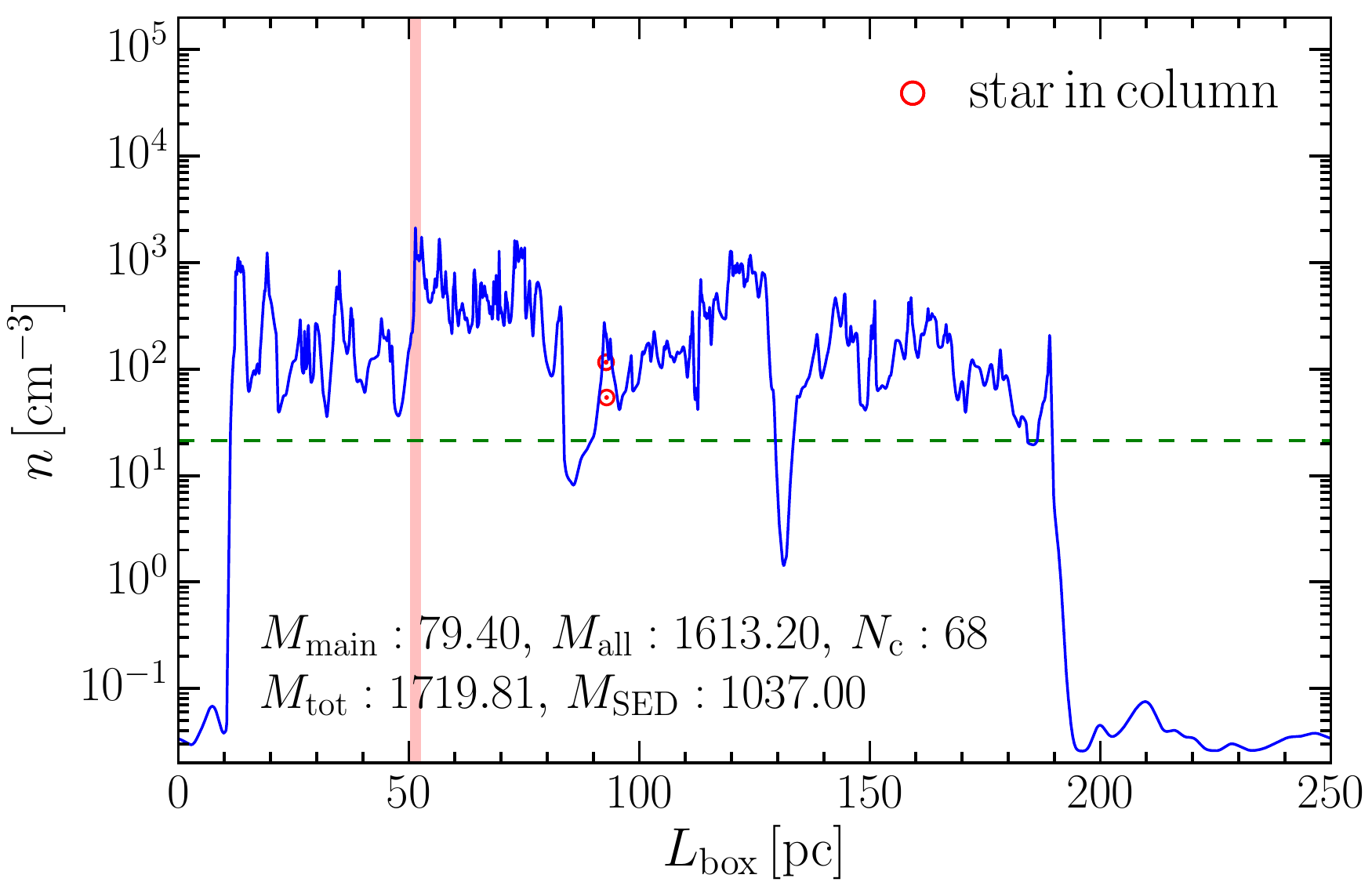}
\caption{Density profile along the line of sight for two example sources (the same two as in Figure~\ref{fig_5band_image}): one source has a dominant clump (left panel) and the other has many clumps of approximately the same peak density (right panel). The red vertical line shows the position of the main clump (the highest density peak), with the thickness of the line corresponding to the source diameter (we assume that the 3D clump has the same diameter as the corresponding source). The open cyan diamonds are the stars located inside the main clump. The open red circles are the stars located in the column of the source footprint, but outside of the dominant clump. The green dashed horizontal line shows the $1\%$ of the maximum density, which is minimum density we have chosen to define secondary clumps.}
\label{fig_density_along_line_of_sight}
\end{figure*}

Projection effects along the line of sight can strongly affect the derivation of observational clump properties, particularly when no kinematic information is available. To establish the importance of projection effects, we search for the three-dimensional (3D) counterparts of the synthetic sources in the simulation datacubes used to generate the synthetic observations. By relating the synthetic sources to their 3D counterparts, we can then compare the observational properties with the intrinsic clump properties.

\subsection{Observational versus Intrinsic Clump Mass} \label{real_mass}

Figure~\ref{fig_density_along_line_of_sight} shows the density profile along the lines of sight of two sources of similar mass (618 and 1037~$\rm M_{\odot}$) at 12~kpc, both classified as protostellar in the synthetic catalog, chosen to be an example of a line of sight dominated by a real 3D clump (left panel) and of a line of sight without a dominant 3D clump (right panel). The gas density is computed within a column of size equal to the source diameter. The open cyan diamonds mark the positions of stars located inside the main 3D clump, while the open red circles are for the stars located in the column of the source footprint, but outside of the main 3D clump. The y-axis coordinate of the diamonds and circles corresponds to the mean density around the star in a $(0.12\,{\rm pc})^3$ volume. While in the example dominated by the main 3D clump (left panel) one star is found inside the main clump, and several more in its vicinity, only two stars are found in the line of sight of the other example (right panel) and at relatively low densities, and none inside the main 3D clump.   

Density profiles like those shown in Figure~\ref{fig_density_along_line_of_sight} are computed for all the sources in our synthetic catalog, and are then used to identify the main 3D clump corresponding to each synthetic source. The main 3D clump in each line of sight is defined as the fraction of the corresponding column centered at the highest density maximum and with a size along the line of sight equal to the synthetic source diameter (the size on the plane of the sky). The position of the maximum density is shown by the vertical transparent line in Figure~\ref{fig_density_along_line_of_sight}, whose thickness corresponds to the diameter of the synthetic source and thus of the main 3D clump. We also define as independent 3D clumps all other density maxima with value larger than 1\% of the absolute maximum that defines the main 3D clump (see the horizontal dashed line). All 3D clumps are assumed to have a size in the direction of the line of sight equal to the source diameter as well. 

With these definitions of main and secondary 3D clumps, we can then associate three different mass values to each line of sight, the total mass of the column, $M_{\rm tot}$, the sum of the masses of all clumps, $M_{\rm all}$, and the mass of the main clump, $M_{\rm main}$, and compare them with the mass of the corresponding synthetic source, $M_{\rm SED}$. In the two examples of Figure~\ref{fig_density_along_line_of_sight}, the main 3D clump in the left panel has a mass $M_{\rm main}=305.7\,\rm M_{\odot}$, approximately half the value of the estimated source mass, $M_{\rm SED}$, while the main 3D clump in the right panel has a mass $M_{\rm main}=79.4\,\rm M_{\odot}$, approximately 13 times smaller than $M_{\rm SED}$. 

\begin{figure}
\centering
\includegraphics[width=0.48\textwidth]{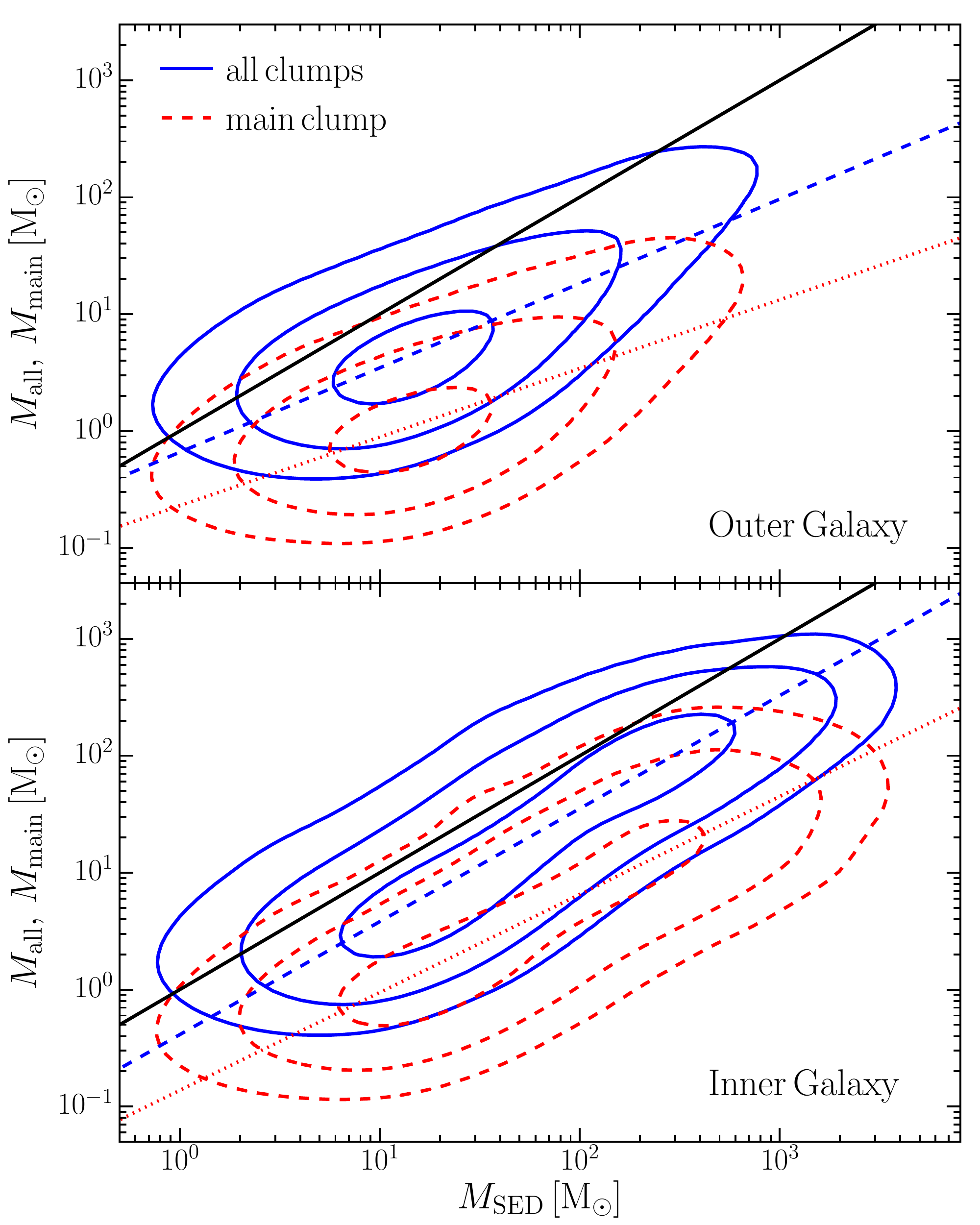}
\caption{Sum of all the 3D clump masses, $M_{\rm all}$ (solid contour lines), and main 3D clump mass, $M_{\rm main}$ (dashed contour lines), versus the source mass derived from the SED fitting, $M_{\rm SED}$, for all the sources in the synthetic catalog, normalized with the Outer-Galaxy distance distribution (upper panel) and the Inner Galaxy distance distribution (lower panel). The solid black line is the one-to-one relation. The dashed straight line is the least-squares fit to the $M_{\rm all}-M_{\rm SED}$ relation, 
$M_{\rm all}=(0.658\pm 0.046)M_{\rm SED}^{(0.722\pm 0.023)}$ for the Outer Galaxy and $M_{\rm all}=(0.412\pm 0.027)M_{\rm SED}^{(0.967\pm 0.015)}$ for the Inner Galaxy case, with the masses in units of $\rm M_{\odot}$. The dotted line is the least-squares fit to the $M_{\rm main}-M_{\rm SED}$ relation,
$M_{\rm main}=(0.230\pm 0.014)M_{\rm SED}^{(0.587\pm 0.020)}$ for the Outer Galaxy
and $M_{\rm main}=(0.137\pm 0.009)M_{\rm SED}^{(0.838\pm 0.015)}$ for the Inner Galaxy case.}
\label{fig_3d_mass}
\end{figure}

The comparison of $M_{\rm all}$ and $M_{\rm main}$ with $M_{\rm SED}$ for all the sources in our catalog is shown in Figure~\ref{fig_3d_mass}, where the black solid line marks the one-to-one relation. The least-squares fits to the median value of each of the two masses as a function of $M_{\rm SED}$ are shown in Figure~\ref{fig_3d_mass} as dashed and dotted lines. The most striking result of this comparison is that $M_{\rm SED}$ overestimates the true mass of the main 3D clump, $M_{\rm main}$, by a factor greater than 10, on average (dashed contour lines and dotted least-squares fit). Because the median relation given by the fit is rather shallow, $M_{\rm main}=(0.230\pm 0.014)M_{\rm SED}^{(0.587\pm 0.020)}$ for the Outer Galaxy and $M_{\rm main}=(0.137\pm 0.009)M_{\rm SED}^{(0.838\pm 0.015)}$ (the masses are in $\rm M_{\odot}$), the ratio $M_{\rm SED}/M_{\rm main}$ increases with increasing $M_{\rm SED}$: the larger the value of $M_{\rm SED}$, the larger the contribution to that mass from structures along the line of sight unrelated to the main 3D clump. In other words, {\it most of the mass of a CuTEx source is due to the overlap of different structures, and increasingly so towards larger masses.} In the case of the Hi-GAL catalog, this effect is likely to be even stronger than in our simulation, as discussed in \S~\ref{caveats}.

Even the sum of all the 3D clumps we have selected along the line of sight, $M_{\rm all}$, is typically smaller than the estimated source mass, with the ratio $M_{\rm SED}/M_{\rm all}\approx 3$ on average. The relation between $M_{\rm tot}$ and $M_{\rm SED}$, $M_{\rm tot}=(2.184\pm 0.129)M_{\rm SED}^{(0.572\pm 0.020)}$ (Outer Galaxy) and $M_{\rm tot}=(1.356\pm 0.076)M_{\rm SED}^{(0.802\pm 0.013)}$ (Inner Galaxy), is also shallower than linear, but $M_{\rm tot}$ is, on average, only slightly smaller than $M_{\rm SED}$, which demonstrates that the SED fitting is able to approximately recover the total mass in the line of sight.

\begin{figure*}
\centering
\includegraphics[width=1.\textwidth]{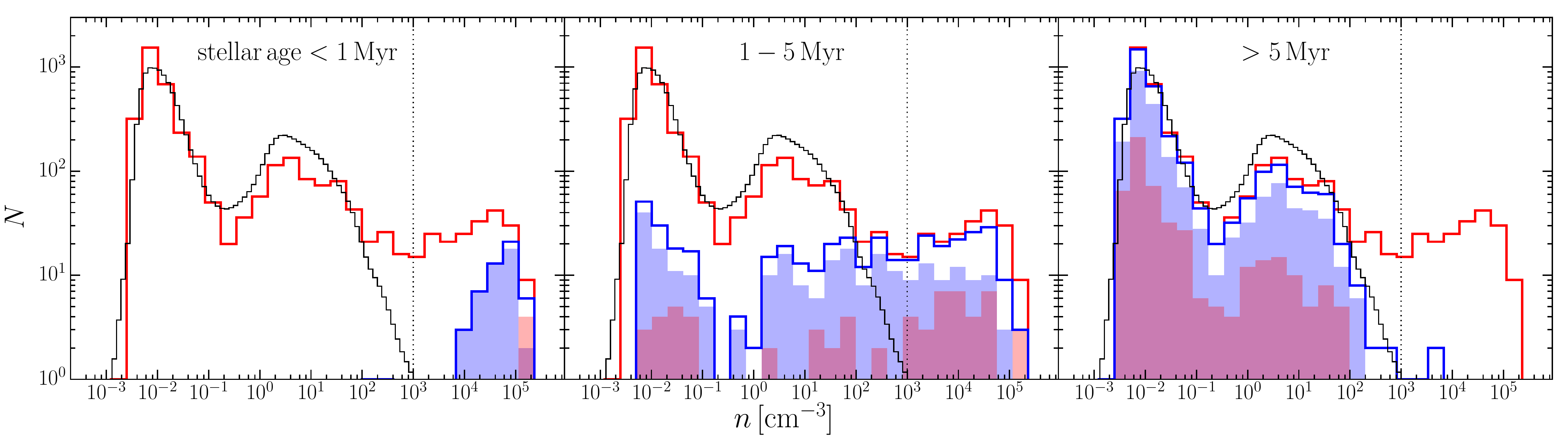}
\caption{Gas density distributions sampled uniformly at a $2048^3$ resolution in the last snapshot of the simulation analyzed in this work, at $t = 34.2$~Myr (solid black line) and sampled at the positions of the stars (solid red line). The density at the position of a star is the local mean gas density calculated in a $\sim 0.12$\,pc box centered around the star. The density around the stars is also plotted as a blue step histogram after separating the stars into three categories based on their age, $<$ 1~Myr, 1-5~Myr and $>$ 5~Myr (left, middle, and right panels respectively). From each of these age intervals, we also show two sub-samples corresponding to stars with mass $<5 \,\rm M_{\odot}$ (blue filled histogram), and with mass $>8 \,\rm M_{\odot}$ (red filled histogram). The overlap region of the two filled histogram is in purple.}
\label{fig_star_mean_density}
\end{figure*}

\begin{figure}
\centering
\includegraphics[width=0.48\textwidth]{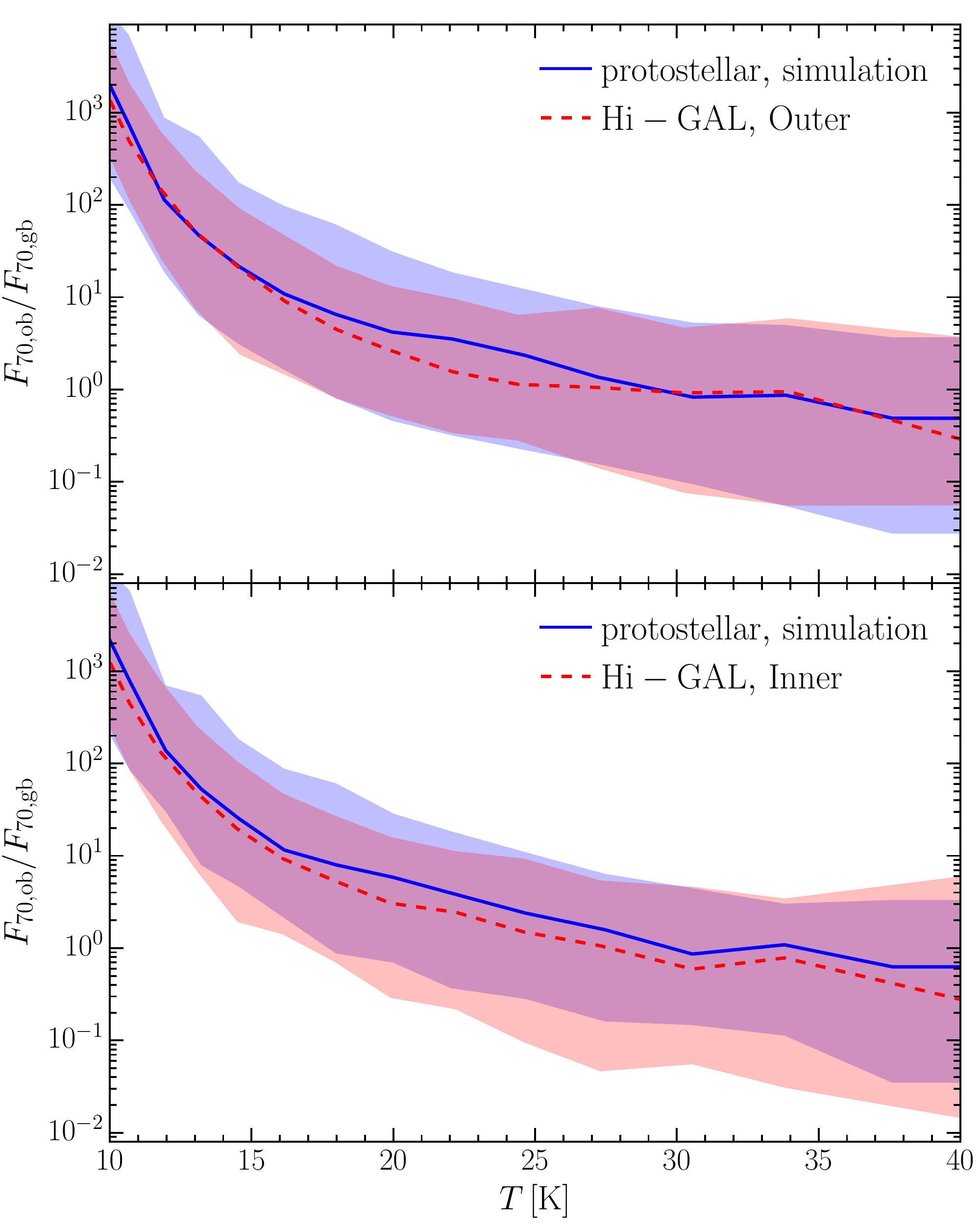}
\caption{The ratio of the observed 70\,$\mu$m flux, $F_{\rm 70,ob}$, and 70\,$\mu$m flux extrapolated from the greybody SED fitting, $F_{\rm 70,gb}$, versus the temperature for the protostellar sources in the synthetic and the Hi-GAL Outer Galaxy catalogs (upper panel), and the Inner Galaxy catalog (lower panel). The solid blue and dashed red lines are the median values for the synthetic protostellar sources and Hi-GAL protostellar sources, respectively. The shaded areas (blue and red, respectively) show the interval between the 2.5 and 97.5 percentiles of the flux ratio.}
\label{fig_f70}
\end{figure}

\begin{figure}
\centering
\includegraphics[width=0.49\textwidth]{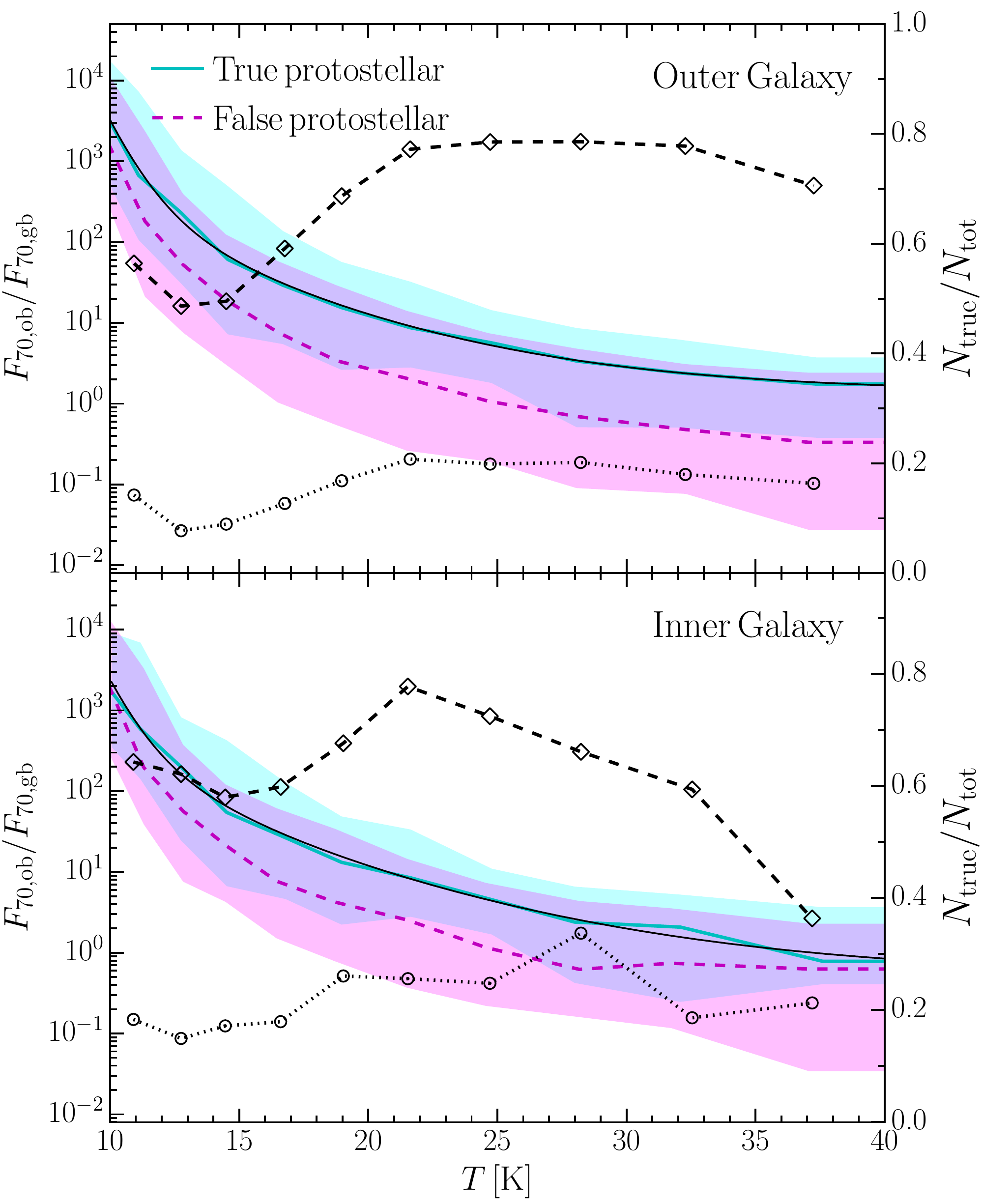}
\caption{The ratio of the observed 70\,$\mu$m flux, $F_{\rm 70,ob}$, and 70\,$\mu$m flux extrapolated from the greybody SED fitting, $F_{\rm 70,gb}$, versus the temperature for the protostellar sources in the synthetic catalog. The cyan line is the median values for the ``true" protostellar sources, which are those with at least one embedded star (with local density $\geq 1000 \rm \, cm^{-3}$) in their line of sight. The magenta line shows the median values for the ``false" protostellar sources, which are those without embedded stars in their line of sight. The shaded areas show the 2.5 and 97.5 percentiles of the 70\,$\mu$m flux ratio. The solid black line is an analytical fit to the cyan median line, and the dashed and dotted black lines are the fraction of ``true" protostellar sources of all sources above and below the solid black line, respectively, with the value of these fractions given by the scale on the right y-axis.}
\label{fig_f70_true}
\end{figure}

\subsection{Starless versus Protostellar Clumps} \label{protostellar}

Having associated our synthetic sources with 3D clumps, we can now ask whether the sources classified as protostellar are associated with actual 3D protostellar clumps or not. For this possible association we consider both the main 3D clump and other secondary clumps along the line of sight of each source. To address this question we need a criterion to discriminate between actual embedded stars and other stars that may be in the line of sight by pure chance, rather than by birth. For that purpose, we adopt a simple density threshold, such that a star is considered to be embedded if the mean density in a $(0.12\,{\rm pc})^3$ cell around it is larger than $10^3\,\rm cm^{-3}$. Among all the protostellar sources in our catalog, we then define as true protostellar sources those with at least one embedded star in their line of sight, and as false protostellar sources the ones without embedded stars, irrespective of their 70\,$\mu$m flux. In the two examples of Figure~\ref{fig_density_along_line_of_sight}, the source from the left panel is a true protostellar clump (it has seven stars at density larger than $10^3\,\rm cm^{-3}$), while the source from the right panel is a false protostellar clump (it has no stars above the density threshold).

The value of the density threshold is justified by a comparison of the gas density distribution over the whole box with the density distribution sampled by the positions of the stars. The solid black line histogram in the three panels of Figure~\ref{fig_star_mean_density} shows the gas density distribution computed from the final simulation snapshot analyzed in this work at a uniform resolution of 2,048$^3$ ($\sim0.12$~pc). The solid red line histogram in the same panels is the gas density distribution sampled at the same resolution, but only at the positions of the stars at the time of the final snapshot (the density is correctly centered at the stellar positions by using a higher-resolution extraction of the density field and then averaging the density around each star within a cell of 0.12~pc). Figure~\ref{fig_star_mean_density} shows that the uniformly-sampled density distribution is bimodal, with the lower density peak at $\sim0.01\,\rm cm^{-3}$ corresponding primarily to hot gas, and the higher density peak at $\sim3\,\rm cm^{-3}$ primarily to colder gas. The density distribution sampled by the stars, instead, has three peaks. The two lower density peaks are the same as for the global distribution, while the highest density one, at $\sim3\times 10^4\,\rm cm^{-3}$, corresponds to a characteristic density of protostellar cores. 

Because protostellar cores occupy a tiny fraction of the computational volume, their density is sampled by the global gas distribution (black line) only at extremely low probability, not shown in the plots of Figure~\ref{fig_star_mean_density}. The densest peak is instead visible in the density distribution sampled by the stars because the fraction of stars that are found within their parent protostellar cores is orders of magnitude in excess of the fraction of the total volume occupied by the cores. The fact that the highest density peak is associated with protostellar cores is confirmed by the comparison of the three panels in Figure~\ref{fig_star_mean_density}, where the density distribution sampled only by stars in a limited age range is shown in each panel by the blue-line histogram. Considering only stars younger than 1~Myr (left panel), one can see that they sample only density around the peak, while the stars older then 5~Myr (right panel) sample only densities outside of that peak. Furthermore, when stars are older (right panel) they sample well the global bimodal density field, meaning that their positions not only do not correspond to dense protostellar cores any longer, but are also completely random with respect to the density field. 

Having identified the highest density peak as due to protostellar cores, the position of the minimum between the two denser peaks in the density distribution sampled by the stars, at $\sim 10^3\,\rm cm^{-3}$, can be adopted as the threshold between the overall density distribution and the density of protostellar cores. This justifies the density threshold used in our criterion to define embedded stars and thus to discriminate between true and false protostellar sources.

\begin{figure*}
\centering
\includegraphics[width=1.\textwidth]{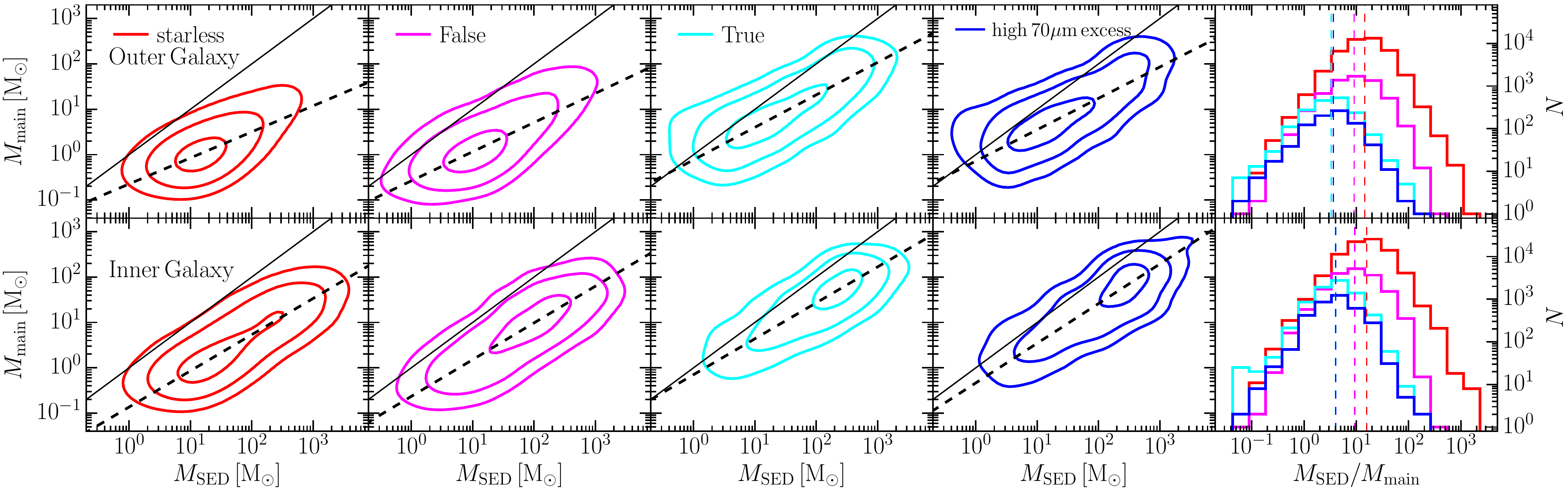}\\
\caption{Main 3D clump mass versus SED mass, as in Fig.~\ref{fig_3d_mass}, for starless, ``false" protostellar, ``true" protostellar, and protostellar sources selected based on the 70~$\mu$m excess (see Fig.~\ref{fig_f70_true}), from left to right. The thin solid black line is the 1-to-1 relationship. The fits shown by the dashed black lines are: $M_{\rm starless}=(0.226\pm 0.012)M_{\rm SED}^{(0.574\pm 0.018)}$, $M_{\rm false}=(0.264\pm 0.018)M_{\rm SED}^{(0.643\pm 0.023)}$, $M_{\rm true}=(0.769\pm 0.057)M_{\rm SED}^{(0.716\pm 0.021)}$, and $M_{\rm high}=(0.722\pm 0.073)M_{\rm SED}^{(0.691\pm 0.030)}$ for the Outer Galaxy. For the Inner Galaxy: $M_{\rm starless}=(0.136\pm 0.008)M_{\rm SED}^{(0.798\pm 0.013)}$, $M_{\rm false}=(0.234\pm 0.018)M_{\rm SED}^{(0.810\pm 0.017)}$, $M_{\rm true}=(0.707\pm 0.095)M_{\rm SED}^{(0.791\pm 0.024)}$, and $M_{\rm high}=(0.519\pm 0.087)M_{\rm SED}^{(0.862\pm 0.029)}$. All the masses in the previous relations are in units of $\rm M_{\odot}$. The rightmost panels show the distributions of the $M_{\rm SED} / M_{\rm main}$ ratio.}
\label{fig_real}
\end{figure*}

We have found that true protostellar sources can be partially differentiated from false protostellar sources based on purely observable quantities, using the dependence of their 70\,$\mu$m excess on their temperature. Before demonstrating this method on the synthetic sources, we want to verify that the relation between 70\,$\mu$m excess and temperature of the synthetic protostellar sources reproduces well that of the observations. The 70\,$\mu$m excess is defined as the ratio of the 70\,$\mu$m flux of the source, $F_{\rm 70,ob}$, and the grey-body 70\,$\mu$m flux, $F_{\rm 70,gb}$, that is the 70\,$\mu$m flux extrapolated from the SED fit of the fluxes at the other wavelengths. The upper panel of Figure~\ref{fig_f70} shows that the temperature dependence of $F_{\rm 70,ob}/F_{\rm 70,gb}$ for the protostellar sources in our synthetic catalog (blue shaded area) has approximately the same median values and dispersion as for the Hi-GAL sources in the Outer Galaxy (red shaded area). Compared with the Inner Galaxy, the 70\,$\mu$m excess of the synthetic sources is slightly larger than for the Hi-GAL sources (lower panel). 

The relation between the 70\,$\mu$m excess and the temperature of false protostellar synthetic sources is shown by the magenta shaded areas in Figure~\ref{fig_f70_true}. The cyan shaded areas corresponds to the true protostellar synthetic sources. In the upper panel, the number of sources at different distances is the same as in the synthetic catalog, to mimic the distance distribution of sources in the Outer Galaxy, while in the lower panel the number of sources has been renormalized according to their distance to mimic the distance distribution in the Inner Galaxy, as in Figure~\ref{fig_diameter}. The dashed red line and the cyan solid line show the median values of $F_{\rm 70,ob}/F_{\rm 70,gb}$ as a function of temperature, measured in logarithmic temperature intervals, for the false and true prestellar sources respectively. The true protostellar sources are systematically warmer, at equal value of $F_{\rm 70,ob}/F_{\rm 70,gb}$, or, equivalently, they have systematically larger $F_{\rm 70,ob}/F_{\rm 70,gb}$ at equal value of temperature (by almost an order of magnitude in the case of the Outer Galaxy's distance distribution shown in the upper panel). However, there is also a considerable overlap between the two types of sources. 

To partially separate true from false protostellar sources, we propose to separate the protostellar sources in two groups using the median relation for the true sources (cyan line). We first obtain an analytical fit to the cyan line for the Outer Galaxy distance distribution, 
\begin{equation}
F_{\rm 70,ob}/F_{\rm 70,gb}= \frac{2877}{(T/10\,{\rm K})^{15}} + \frac{367}{(T/10\,{\rm K})^5} + 1.3  \,,
\label{fit_outer}
\end{equation}
and for the Inner Galaxy case,
\begin{equation}
F_{\rm 70,ob}/F_{\rm 70,gb}= \frac{2051}{(T/{\rm 10\,K})^{15}} + \frac{365}{(T/{\rm 10\,K})^5} + 0.5  \,,
\label{fit_inner}
\end{equation}
which is shown by the solid black lines in Figure~\ref{fig_f70}. The dashed black lines connecting the diamond symbols show the ratio between true protostellar sources and total protostellar sources found above the analytical fits. The values of this ratio are shown in the y-axis on the right-hand side of the plots. The lower black dotted lines, connecting the circle symbols, show the same ratio for sources below the analytical fit to the median line. One can see that, in the case of the Outer Galaxy, approximately 80\% of the protostellar sources are true ones in the approximate temperature range between 21~K and 33~K, while below the lines the fraction of true protostellar sources is always lower than 20\%. Averaging over all temperatures, the fraction above the line is 0.64 and 0.66 for the Outer and Inner Galaxy respectively, while the fraction below the line is 0.14 and 0.21, for the same cases. In summary, the method proposed here allows to select a class of sources where 60-70\% are true protostellar ones (they have real 3D clumps along the line of sight containing embedded protostars), by selecting half of the total number of candidate protostellar sources. 

Because we can partially extract true protostellar sources, it is interesting to check the relation of their estimated mass, $M_{\rm SED}$, with the mass of the corresponding main 3D clumps, $M_{\rm main}$, for this specific subset of sources. Figure~\ref{fig_real} shows contour plots of the relation between $M_{\rm main}$ and $M_{\rm SED}$ for the synthetic sources classified as starless (left panel), false protostellar (second panel from the left), true protostellar (third panel from the left), and for the sources above the line defined by Equation~\ref{fit_outer} (fourth panel from the left). While for false protostellar sources the mass discrepancy, $M_{\rm SED} \gg M_{\rm main}$, is comparable to that of starless sources, for true protostellar sources it is significantly reduced. The fifth panel of Figure~\ref{fig_real} shows the probability distributions of the ratio $M_{\rm SED} / M_{\rm main}$ for the four populations, with the vertical dashed lines corresponding to the median values. The median values decrease from nearly 20 for starless sources to $\approx 4$ for true protostellar ones.   

Even sources selected with the observational criterion suggested above, based on the relation in Equation~\ref{fit_outer}, have a median ratio $M_{\rm SED} / M_{\rm main}\approx 4$. This analysis shows that, a source selection that gives a majority of true protostellar sources also gives sources that are less affected by (though not free from) projection effects.

\section{Discussion} \label{discussion}

\subsection{Implications for Clump Dynamics and Evolution} \label{bound}

We have found that the estimated mass of our synthetic sources, $M_{\rm SED}$, is typically of the order of the total mass in the line of sight to the source, and an order of magnitude larger than the densest real 3D clump identified in the simulation along the line of sight. This implies that our synthetic sources, and by extension the Hi-GAL sources, should not be considered as individual clumps. If interpreted as individual clumps, one should keep in mind that the masses of such clumps are most likely overestimated by a very large factor, which complicates the analysis of their dynamical state (e.g. bound versus unbound \citep{Elia+2017}), their evolutionary state (e.g. the clump mass-luminosity relation \citep{Molinari+2008}), their statistical properties (e.g. the velocity-size relation \citep{Traficante+18}), and their role in the formation of massive stars (e.g. the estimated infall rates \citep{Traficante+18}, or the estimated column density threshold \citep{Tan+2014,Urquhart+2014}). Even the selection of a subset of sources with molecular emission-line spectra without multiple components \citep[e.g.][]{Traficante+18} may not be sufficient to prevent projection effects, as the existence of a large number of unrelated sources along the line of sight could presumably result in the appearance of an approximately Gaussian velocity profile.      

In\S~\ref{correlations} we found that our synthetic sources, as well as the Hi-GAL sources, follow very closely Larson's mass-size relation, corresponding to their surface density being nearly independent of size, on average. \citet{Elia+2017} implicitly assumed that this mass-size relation of Hi-GAL sources had to stem from the combination of Larson's velocity-size and velocity-mass relations \citep{Larson81}, so that sources above (more massive than) the average mass-size relation could be interpreted as gravitationally bound, and sources below (less massive than) the relation would be unbound. In view of our finding that $M_{\rm SED}$ is not a reliable estimate of a real clump mass, the interpretation of the mass-size relation of Hi-GAL sources is not straightforward. In fact, it was later found that Hi-GAL sources do not follow Larson's velocity-size relation at all \citep{Traficante+18}, which is in itself a result of difficult interpretation if the velocity dispersion has potentially multiple contributions along the line of sight (not necessarily spotted as multiple spectral components), but certainly invalidates the bound versus unbound classification of Hi-GAL sources based on the mass-size relation.

Observed infall rates of massive clumps \citep[e.g.][]{Fuller+05,Beuther+12,Peretto+13,Beuther+13,Wyrowski+16,Traficante+17,Traficante+18,Contreras+18,Yuan+18} are often used to constrain the formation timescale of massive stars. Besides the difficulty of interpreting the kinematic information from emission line spectra in terms of infall, and the need to explain to low detection rate of infall signatures in massive clumps, it is important to remember that a massive clump may host the formation of multiple stars. Furthermore, in view of the results of this study, the clump mass may be grossly overestimated. The infall rate is usually estimated assuming that the infalling gas has a density equal to the clump's mean density. Thus, if the clump mass is overestimate by a factor of 10, the infall rate is also overestimated by the same factor. 

\begin{figure}
\centering
\includegraphics[width=0.48\textwidth]{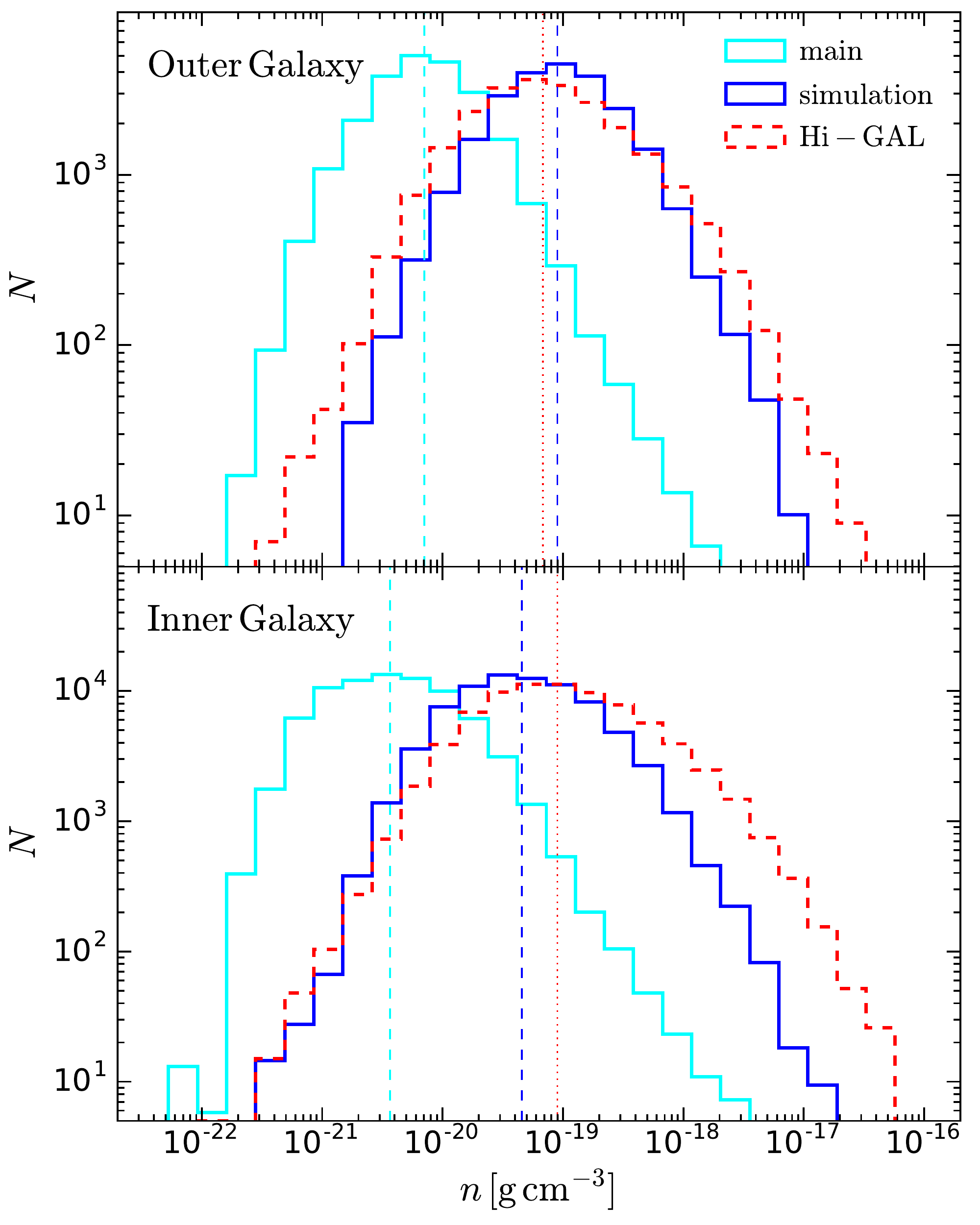}
\caption{Distributions of the mean gas density of the sources in the synthetic catalog (solid blue lines) and in the Hi-GAL catalog with distances between 1.5 and 13.5~kpc (red dashed lines), for the Outer Galaxy (upper panel) and the Inner Galaxy (lower panel). In the lower panel, the numbers of synthetic sources at distances $>2$~kpc have been rescaled as in previous figures. The cyan lines show the distributions of the mean density of the main 3D clumps in the lines of sight to the synthetic sources.}
\label{fig_surface_density}
\end{figure}

Figure~\ref{fig_surface_density} shows the distributions of the mean volume density of our synthetic sources (blue histograms) and of the Hi-GAL sources (red dashed line histograms) for the Outer Galaxy (upper panel) and the Lower Galaxy (lower panel). The cyan histograms show the distribution of the mean density of the main 3D clump associated to each of our synthetic sources. This distribution is shifted to lower densities by a factor of 10 relative to that of the synthetic sources, as expected from the mass comparison in \S~\ref{real_mass}. In the case of the Hi-GAL sources, the factor may be even larger, as discussed below in \S~\ref{caveats}. Based on the median values shown by the vertical lines in Figure~\ref{fig_surface_density}, infall rates based on estimated densities of Hi-GAL Inner Galaxy sources may be overestimated by more than a factor of $20$, since the median density in the Hi-GAL Inner Glalaxy catalog is more than a factor of two larger than in our synthetic catalog, which is more than a factor of 10 larger than the median for the corresponding 3D clumps. This uncertainty could be reduced, by approximately a factor of three, if true protostellar sources were selected according to the method suggested in \S~\ref{protostellar}, but even in that case over 30\% of sources would still be false protostellar sources suffering a larger projection effect.

\subsection{From the Simulation Volume to the Galactic Plane} \label{caveats}

\begin{figure}
\centering
\includegraphics[width=0.48\textwidth]{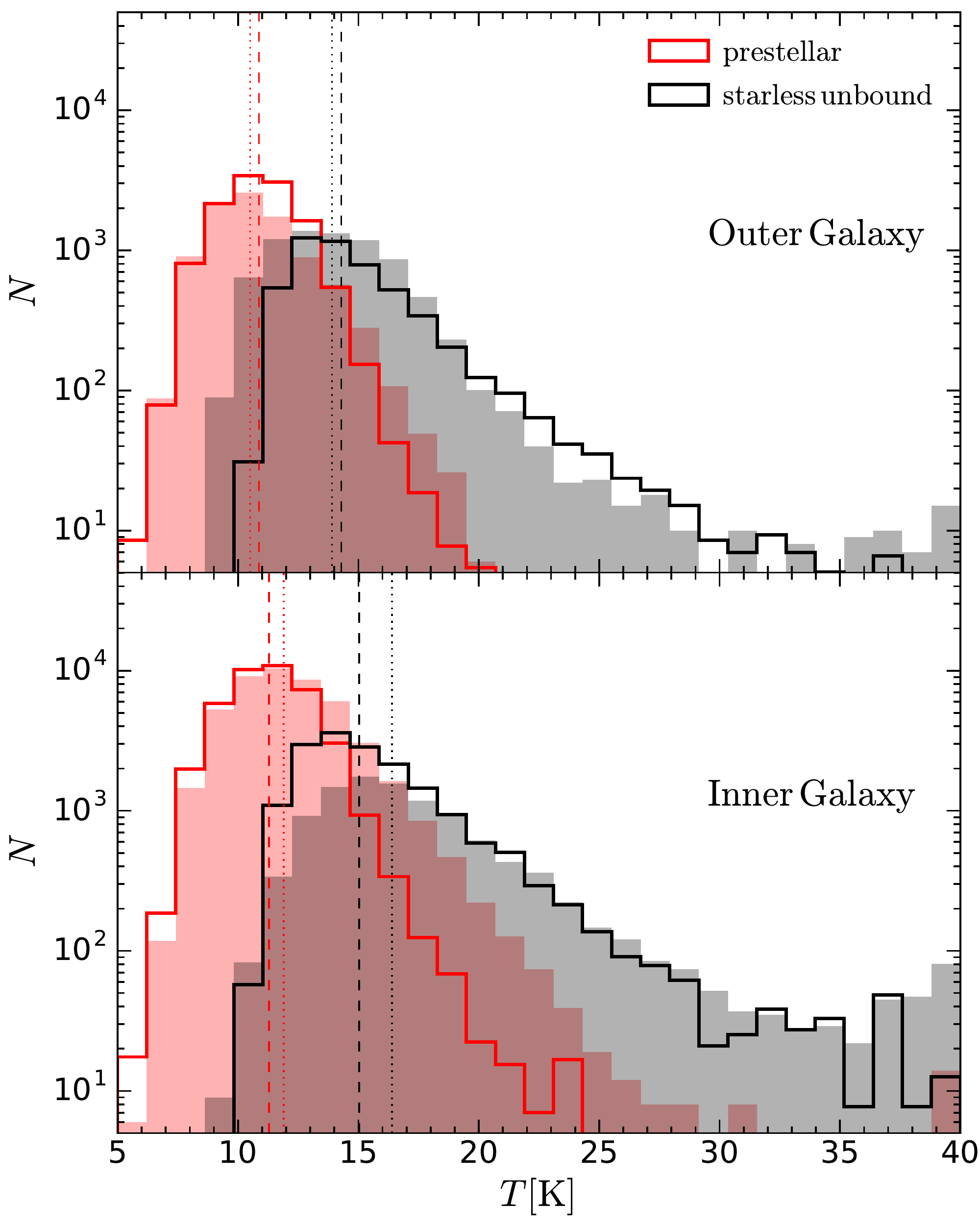}
\caption{Greybody temperature distributions for the synthetic prestellar sources (red line) and the synthetic starless unbound sources (black line), compared with Hi-GAL sources in the distance interval between 1.5 and 13.5~kpc (shaded histograms), for the Outer Galaxy (upper panel) and the Inner Galaxy (lower panel). The dashed and dotted vertical lines are the median values of the synthetic and Hi-GAL sources, respectively. }
\label{fig_temperature_pre}
\end{figure}

The synthetic observations of our simulation reproduce the main observational properties, and their statistical distributions, of the Hi-GAL catalog, with the caveat of a systematic shift of the mass distribution towards lower masses. The mass discrepancy is significant only in the high-mass tail of the protostellar mass distribution in the case of the Outer-Galaxy comparison, while the mass distributions of both starless and protostellar sources in the Inner Galaxy are clearly shifted towards larger masses in both high and low-mass tails, by a factor of $\approx 4$ in mass relative to our synthetic catalog (see Figure~\ref{fig_diameter}). This discrepancy is to be expected, because the 250~pc depth of the computational volume cannot match the full complexity and the highest column densities of the lines of sight toward the Inner Galactic plane sampled by the Hi-GAL survey. 

However, the larger column density available does not mean that the masses of Hi-GAL's sources are more real (closer to the mass of actual 3D clumps) than those in our synthetic catalog. As shown in \S~\ref{real_mass}, the estimated source masses are in large part the result of a projection effect. Thus, we interpret the larger Hi-GAL masses as the result of stronger projection effects in the observations than in the simulation. This is supported by the fact that the mass discrepancy is significantly stronger in the comparison with the Inner Galaxy than with the Outer Galaxy.

Besides the lower values of maximum column densities, the synthetic observations also lack sources of confusion both in the line-of-sight and on the plane of the sky that afflict Galactic disk surveys. Our synthetic observations should be more sensitive to fainter sources, as they are not limited by the confusion due to the Galactic structures, including the cirrus clouds. These cirrus clouds are real objects, but from an observational standpoint, their effect is similar to noise when trying to detect compact sources, and dominates over the purely instrumental noise for wavelengths longer than 70~$\mu$m, as shown in Figure 3 of \citet{Molinari+2016}. Our simulation lacks the equivalent of this cirrus noise, as the AMR has too low resolution of intermediate and low density structures, smoothing them away, and the line-of-sight depth would not be enough to accumulate enough surface brightness to cause comparable confusion.  

In\S~\ref{bound} we have argued that the subdivision of Hi-GAL's starless sources into bound prestellar and unbound ones based on Larson's mass-size relation is most likely incorrect, because of the uncertainty of the source masses demonstrated in this work (see \S~\ref{real_mass}) and because Hi-GAL's sources do not follow Larson's velocity-size relation \citep{Traficante+18}. However, the separation of the sources into these two classes, essentially a cut at nearly constant surface density, offers an additional comparison test between the synthetic catalog and Hi-GAL. For simplicity, we refer to these sources as prestellar and starless unbound, as in \citet{Elia+2017}, even if these names do not reflect their real nature. The only significant discrepancy between our catalog and the observations, with respect to this column density cut, is the temperature distribution, which we show in Figure~\ref{fig_temperature_pre} for the Outer and Inner Galaxy catalogs in the upper and lower panels, respectively.

The temperature distributions of the synthetic sources reproduce reasonably well the observations, except for a significantly deficient tail of high-temperature prestellar sources relative to the Inner Galaxy. The Hi-GAL Inner Galaxy catalog contains an excess of high column density sources (classified as bound prestellar sources) with relatively high temperature, between $\approx 15$~K and $\approx 30$~K, that cannot be explained based on our radiative transfer calculations. The most straightforward explanation is that most of the warmest sources at relatively high column densities are a projection of a number of (starless unbound) sources of low enough column density that the derived temperature can be understood. This projection effect happens also in the synthetic observations, but it is clearly much stronger for the Inner-Galaxy sources. This interpretation is consistent with the fact that the temperature discrepancy of prestellar sources is insignificant in the case of the Outer Galaxy catalog (upper panel of Figure~\ref{fig_temperature_pre}), where projection effects are expected to be significantly reduced, due to the nearly exponential drop in the number of sources as a function of distance beyond $\approx 2$~kpc.

\begin{figure}
\centering
\includegraphics[width=0.23\textwidth]{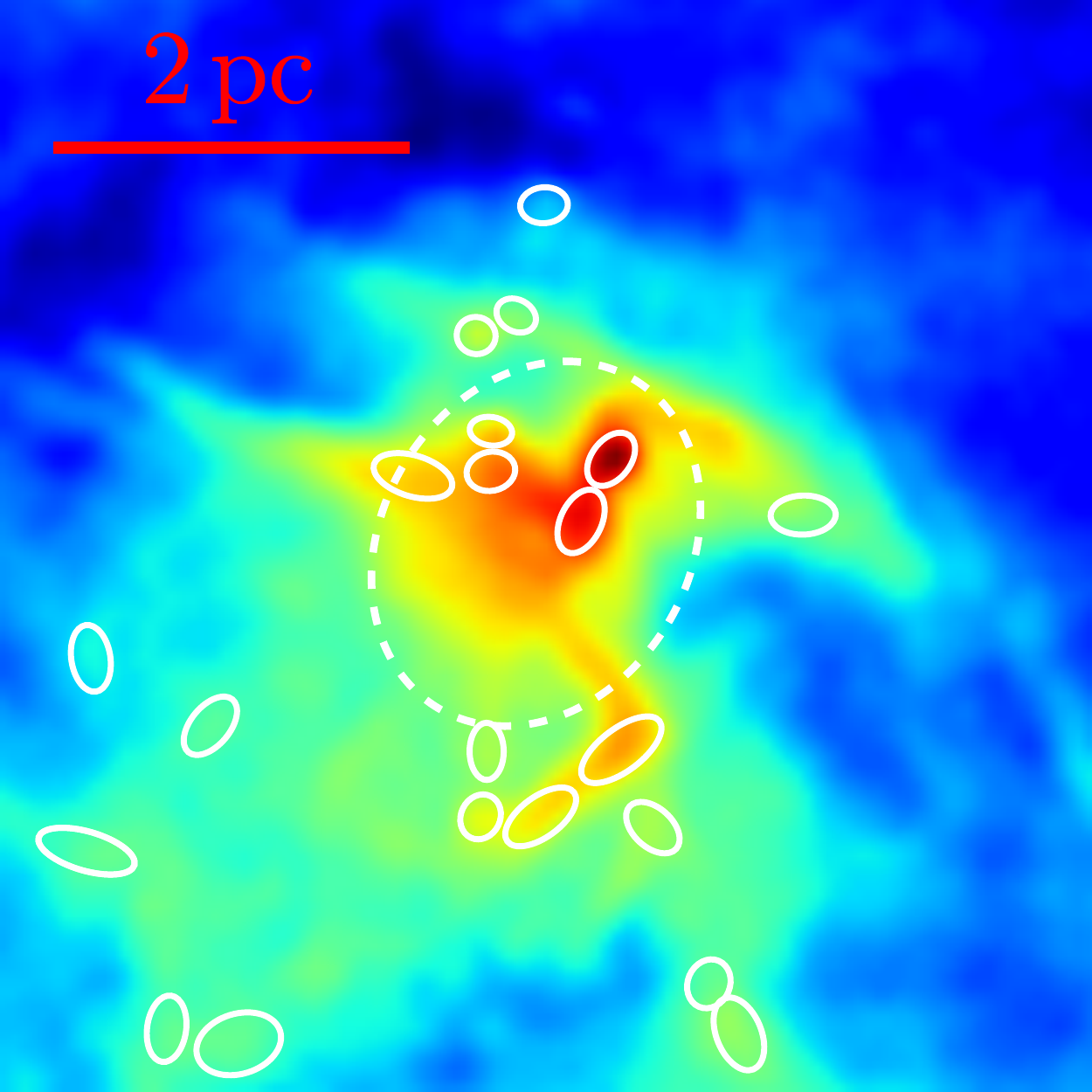}
\includegraphics[width=0.23\textwidth]{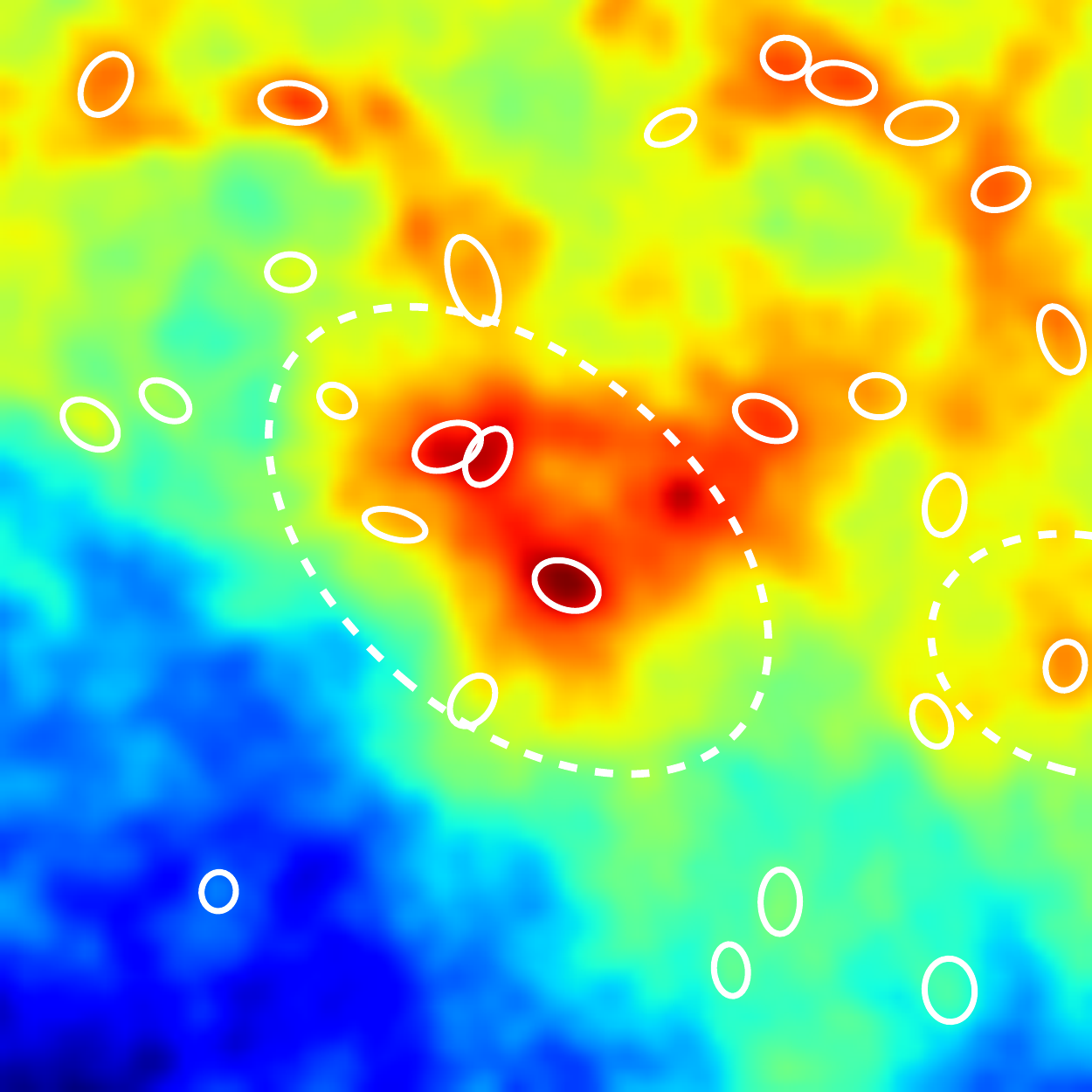}\\
\includegraphics[width=0.23\textwidth]{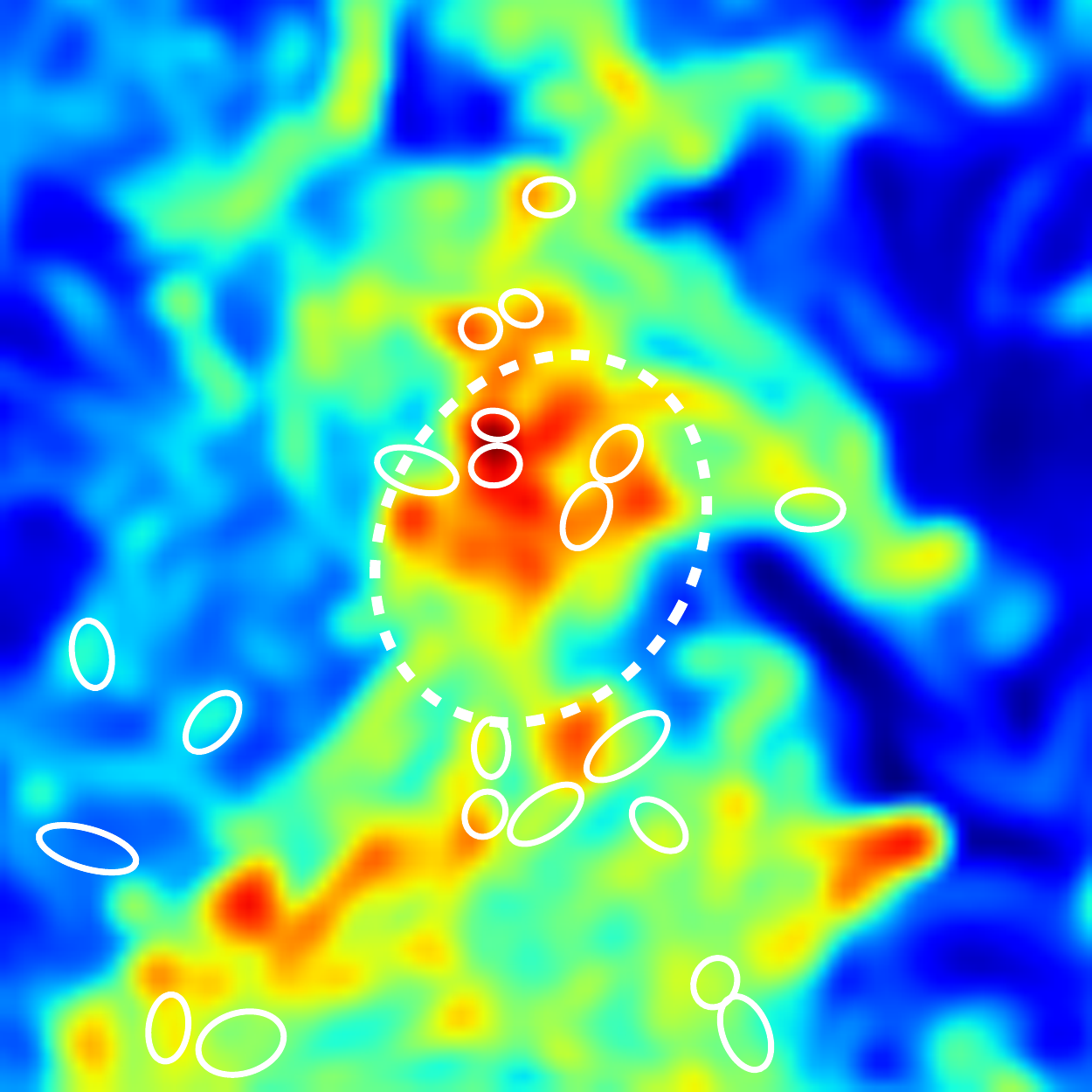}
\includegraphics[width=0.23\textwidth]{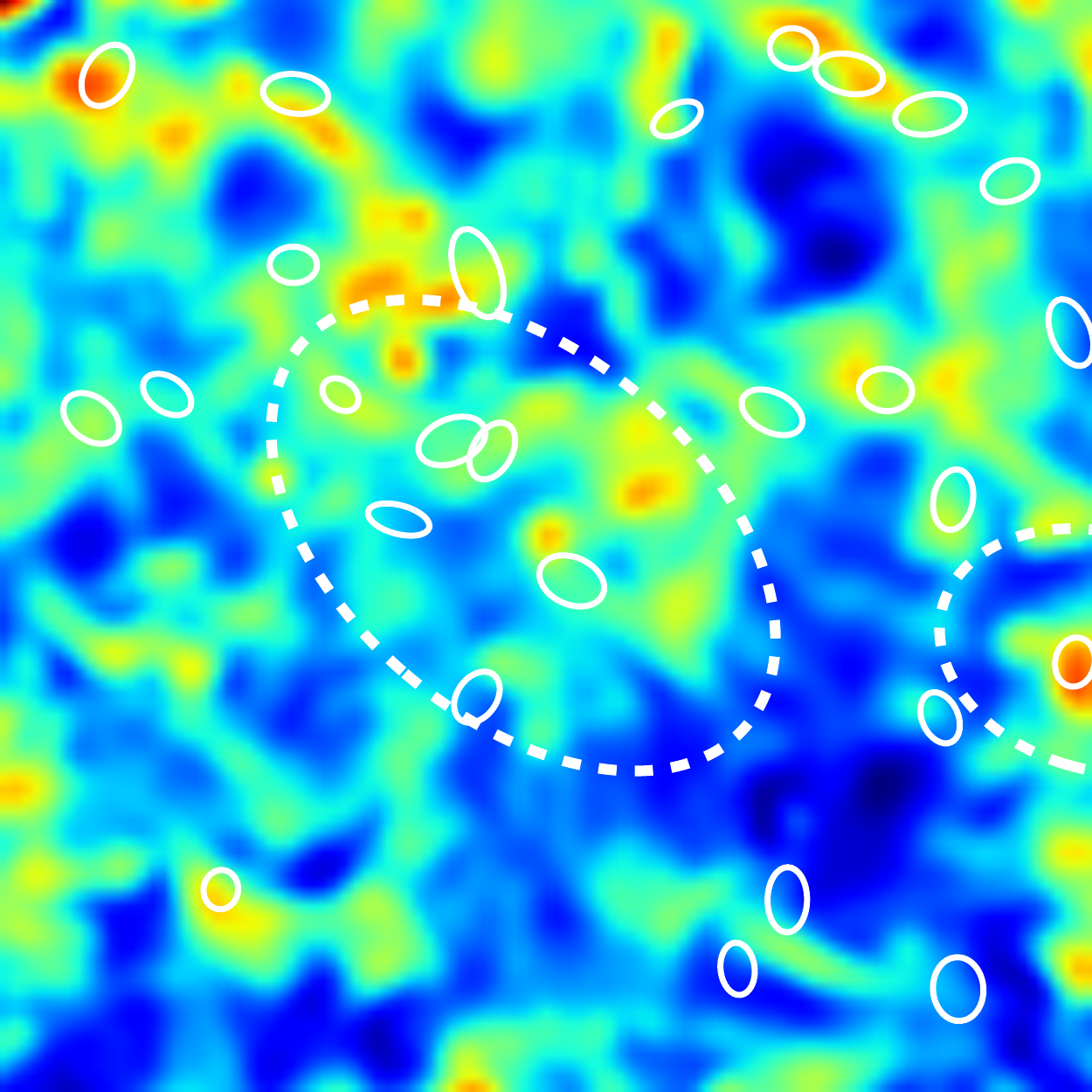}\\
\includegraphics[width=0.23\textwidth]{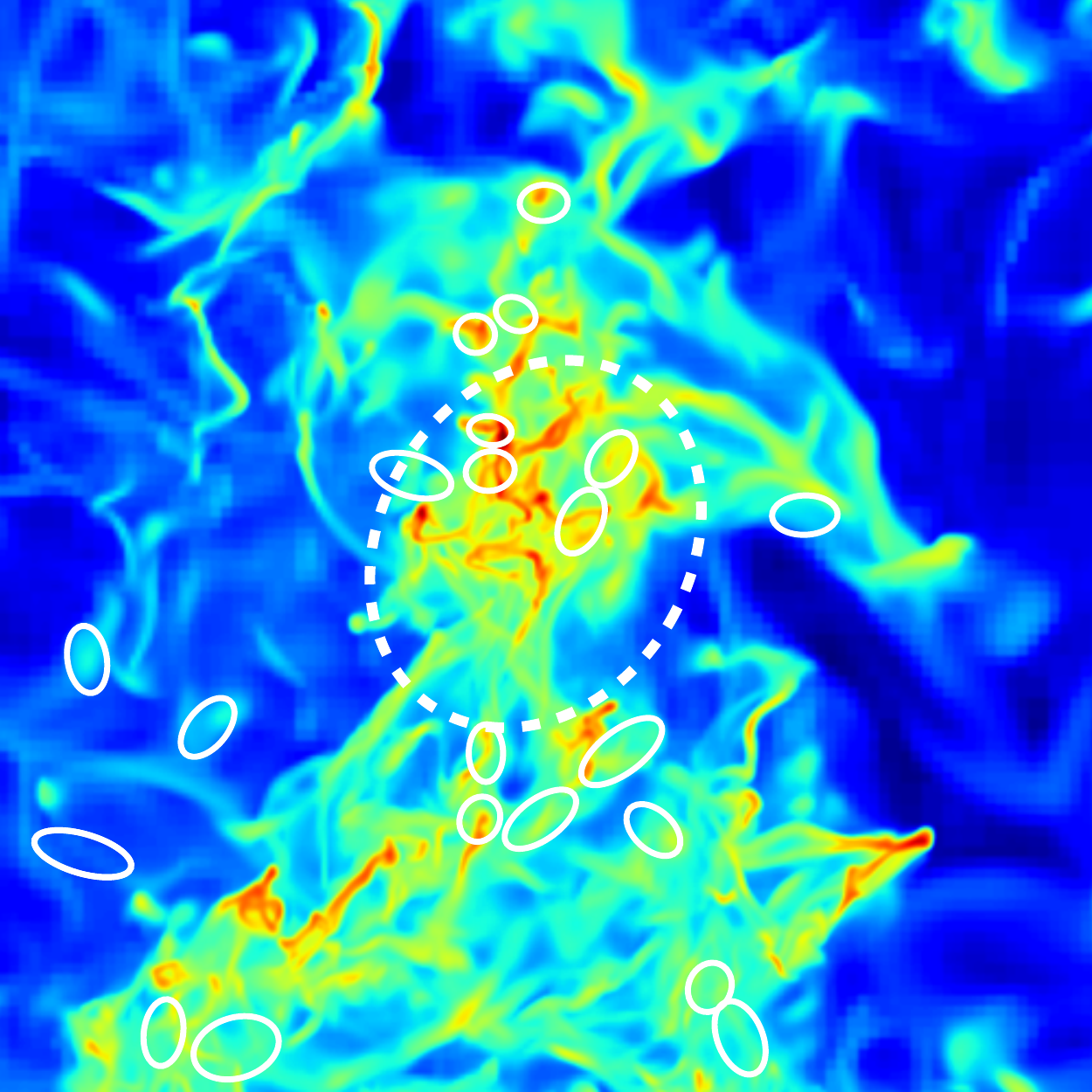}
\includegraphics[width=0.23\textwidth]{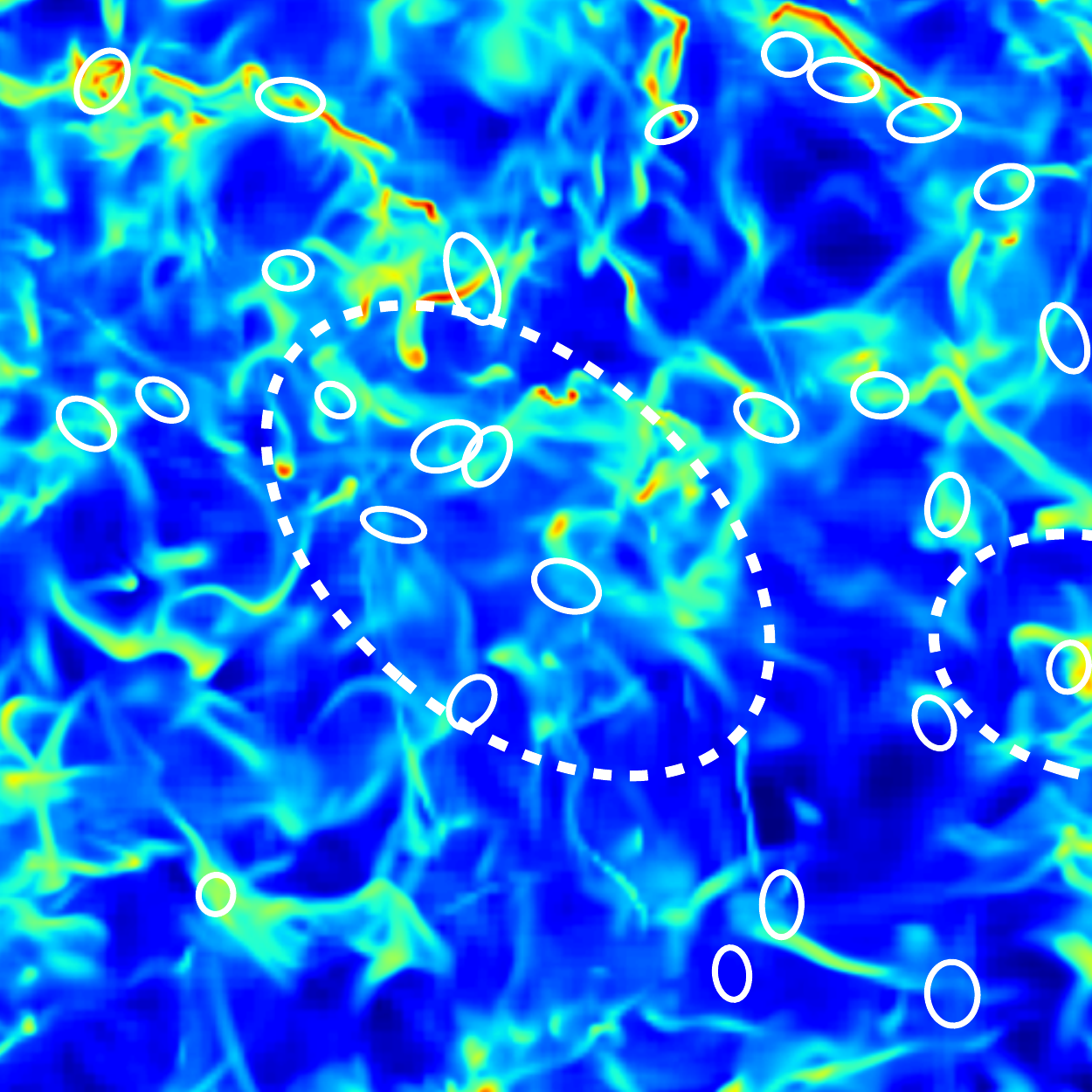}
\caption{The two sources at 12~kpc from Fig.~\ref{fig_5band_image} are shown by the white dashed ellipses, in the left and right columns. The solid white ellipses are the sources found in the same 6~pc regions assumed to be at 2\,kpc. The top panels show the 250~$\mu$m surface brightness maps at the distance of 2~kpc, while the middle panels show the column density at a comparable resolution ($4,096^2$). The lower panels show again the column density in the same regions, but at the highest resolution ($32,768^2$). All the maps have a logarithmic colour scale.}
\label{fig_density}
\end{figure}

\subsection{Angular Resolution and True Nature of Clumps}

So far, we have focused on the uncertainty (mostly the overestimate) of the observed source masses due to blending of different density structures along the line of sight. We have shown that this uncertainty is large (typically a factor of 10) in our synthetic observations (\S~\ref{real_mass}) and thus in the Hi-GAL catalog, as the synthetic observations were obtained by following closely the Hi-GAL data-analysis pipeline. We have further argued that projection effects are likely to be even worse in the Hi-GAL Inner Galaxy catalog, due to the larger column density of lines of sight through the Galactic plane than through our 250~pc volume (\S~\ref{caveats}). Besides these projection effects, the limited spatial resolution of the observations is another important factor that undermines the interpretation of compact sources as individual clumps. At the characteristic distance of Hi-GAL sources, primarily in the approximate range between 2 and 14~kpc, most of the compact sources at the angular resolution of Herschel's observations are expected to be highly fragmented. The effect of the angular resolution (or distance to the source) on the mass determination was already quantified in \citet{Padoan+20massive}, where it was shown that the mass of true prestellar cores observed by Herschel at distances larger than 1~kpc would on average be overestimated by more than a factor of 10. The error grows with increasing distance, being a factor of $\sim 40$ at a distance of $\sim 2$~kpc \citep[see Figure~28 in][]{Padoan+20massive}.

It is generally understood that compact sources from Herschel's observations do not represent individual prestellar cores, but are more likely to be the sites of formation of multiple stars. Their large internal velocity dispersions, revealed by follow-up studies, also implies supersonic turbulence, hence fragmentation, inside the clumps. Nevertheless, unresolved clumps are still often interpreted and modeled as well-defined individual objects within the boundaries of their estimated (and unresolved) size. Rather than attempting a quantitative analysis of specific uncertainties that arise from the limited angular resolution of the observations, we use our synthetic observations to describe the problem graphically through two examples. We consider the same two example sources of Figures~\ref{fig_5band_image} and \ref{fig_density_along_line_of_sight}, and show their maps at the same physical scale ($\sim 5$~pc) as in the panels of Figure~\ref{fig_5band_image}. While in that figure the sources were assumed to be at a distance of 12~kpc, in the top panels of Figure~\ref{fig_density} they are assumed to be at a distance of 2~kpc. At this higher spatial resolution, the 250~$\mu$m sources now appear to be connected to filaments at larger scales, and the original sources have broken into five sources (left panel) and six sources (right panel). The middle panels show that the structure of the column density maps, at the same resolution of Herschel's 250~$\mu$m beam, is even more fragmented than that of the dust emission maps. Finally, the bottom panels  show the column density of the same regions at the maximum resolution of the simulation, 0.008~pc. While the source associated with a dominant 3D clump (a true protostellar source) appears to be like a complex conglomerate of dense cores and filaments (left panel), the source that did not have a dominant 3D clump (a false prootostellar source) has turned into a looser association of filaments and cores that barely stands out of the background. 

These examples show that the limited angular resolution may easily lead to masses of compact sources many times larger than the actual cores and filaments they contain, as previously quantified in \citet{Padoan+20massive}. It is important to appreciate that this effect is conceptually distinct from the blending along the line of sight, and so it may further contribute to the mass uncertainty. However, as the spatial resolution of the observations is increased and more fragments are resolved on the plane of the sky, the chance of blending of these smaller structures along the line of sight is decreased, so the mass estimates become more accurate, as demonstrated with synthetic higher angular resolution ALMA observations in \citet{Padoan+20massive}.

\section{Conclusions} \label{conclusions}
We have used synthetic Herschel observations of a star-formation simulation on a scale of 250 pc to generate a catalog of compact sources, with a range of distances between 2 and 12~kpc. The sources have been selected from the synthetic observations with the CuTEx code, following the same procedure as in the compilation of the Hi-GAL compact source catalog. Our synthetic catalog contains 52,543 compact sources and is an invaluable tool to interpret the nature of the 150,223 Herschel's compact sources in the Hi-GAL catalog. This work serves both as a validation of the synthetic observations and as a first interpretation of the nature of the Hi-GAL sources.

To validate the synthetic observations, we have compared statistical distributions and correlations of size, mass and temperature of the synthetic sources with those of Herschel's sources from the Outer and Inner Galaxy catalogs separately. We have found a good agreement with the observations, particularly with respect to the Outer Galaxy catalog, and some discrepancies (stronger in the comparison with the Inner Galaxy) that can be understood considering the limitations of the simulation (smaller depth than the Galactic plane and incomplete stellar IMF). We have then investigated the nature of the selected sources by searching for their counterparts in the three-dimensional data-cubes of the simulations. Our main results are listed in the following.

\begin{itemize}
    \item The source masses overestimate the clump masses by an order of magnitude on average, due to line-of-sight projection. The estimated mass roughly corresponds to the whole mass along the line of sight, while the most massive clump in the line of sight usually contains only about one tenth of the total mass on average.  
    \item A large fraction of sources classified as protostellar are likely to be starless at all values of temperature and 70~$\mu$m excess, as the 70~$\mu$m excess may be caused by stellar sources outside dense clumps.
    \item We have proposed a method to partially discriminate between false and true protostellar sources based on the depednence of the 70~$\mu$m excess on the temperature. 
    \item We have found evidence of significantly stronger projection effects in the Inner Galaxy catalog than in the Outer Galaxy and synthetic catalogs, from the mass distribution of the sources and their temperature distributions above and below Larson's mass-size relation. This would suggest that the mass of Hi-GAL sources in the Inner Galaxy may be on average over 20 times larger than the main 3D clumps in their lines of sight.
    \item At higher angular resolution, most Hi-GAL clumps should reveal a strongly fragmented structure, so future ALMA observations should confirm the important role of the projection effects demonstrated in this work.
\end{itemize}

Our synthetic catalog will be used to interpret the results of follow-up studies of Hi-GAL sources, including single-dish or higher-resolution ALMA observations of molecular emission lines. In a future work, we will focus on the interpretation of the observational estimates of infall rates of massive clumps, as these have direct consequences for our understanding of the origin of massive stars. We can already conclude from the results of this work that the observed infall rates of massive clumps may have been overestimated by more than one order of magnitude, as the derived masses of Hi-GAL sources cannot be interpreted as the masses of individual clumps. The implications of our results for the formation of massive stars should also be addressed in future works.

\section*{Acknowledgements}
ZJL acknowledges financial support from China Scholarship Council (CSC) under grant No. 201606660003.
PP and VMP acknowledge support by the Spanish MINECO under project AYA2017-88754-P. 
Computing resources for this work were provided by the NASA High-End Computing (HEC) Program through the NASA Advanced Supercomputing (NAS) Division at Ames Research Center. We acknowledge PRACE for awarding us access to Curie at GENCI@CEA, France. Storage and computing resources at the University of Copenhagen HPC centre, funded in part by Villum Fonden (VKR023406), were used to carry out part of the data analysis.

\section*{Data Availability}

The surface brightness, column density, and the full synthetic catalog (the description of all the column can be found in Appendix \ref{appA}), as well as other supplemental material, can also be obtained from a dedicated public URL (\url{ ??? }).

\bibliographystyle{mnras}
\bibliography{massive_clump}

\begin{thebibliography}{}
\makeatletter
\relax
\def\mn@urlcharsother{\let\do\@makeother \do\$\do\&\do\#\do\^\do\_\do\%\do\~}
\def\mn@doi{\begingroup\mn@urlcharsother \@ifnextchar [ {\mn@doi@}
  {\mn@doi@[]}}
\def\mn@doi@[#1]#2{\def\@tempa{#1}\ifx\@tempa\@empty \href
  {http://dx.doi.org/#2} {doi:#2}\else \href {http://dx.doi.org/#2} {#1}\fi
  \endgroup}
\def\mn@eprint#1#2{\mn@eprint@#1:#2::\@nil}
\def\mn@eprint@arXiv#1{\href {http://arxiv.org/abs/#1} {{\tt arXiv:#1}}}
\def\mn@eprint@dblp#1{\href {http://dblp.uni-trier.de/rec/bibtex/#1.xml}
  {dblp:#1}}
\def\mn@eprint@#1:#2:#3:#4\@nil{\def\@tempa {#1}\def\@tempb {#2}\def\@tempc
  {#3}\ifx \@tempc \@empty \let \@tempc \@tempb \let \@tempb \@tempa \fi \ifx
  \@tempb \@empty \def\@tempb {arXiv}\fi \@ifundefined
  {mn@eprint@\@tempb}{\@tempb:\@tempc}{\expandafter \expandafter \csname
  mn@eprint@\@tempb\endcsname \expandafter{\@tempc}}}

\bibitem[\protect\citeauthoryear{{Beuther}, {Linz}  \& {Henning}}{{Beuther}
  et~al.}{2012}]{Beuther+12}
{Beuther} H.,  {Linz} H.,   {Henning} T.,  2012, \mn@doi [\aap]
  {10.1051/0004-6361/201219128}, \href
  {https://ui.adsabs.harvard.edu/abs/2012A&A...543A..88B} {543, A88}

\bibitem[\protect\citeauthoryear{{Beuther}, {Linz}  \& {Henning}}{{Beuther}
  et~al.}{2013}]{Beuther+13}
{Beuther} H.,  {Linz} H.,   {Henning} T.,  2013, \mn@doi [\aap]
  {10.1051/0004-6361/201321498}, \href
  {https://ui.adsabs.harvard.edu/abs/2013A&A...558A..81B} {558, A81}

\bibitem[\protect\citeauthoryear{{Compi{\`e}gne} et~al.,}{{Compi{\`e}gne}
  et~al.}{2011}]{Compiegne_2011}
{Compi{\`e}gne} M.,  et~al., 2011, \mn@doi [\aap]
  {10.1051/0004-6361/201015292}, \href
  {https://ui.adsabs.harvard.edu/abs/2011A&A...525A.103C} {525, A103}

\bibitem[\protect\citeauthoryear{{Contreras} et~al.,}{{Contreras}
  et~al.}{2018}]{Contreras+18}
{Contreras} Y.,  et~al., 2018, \mn@doi [\apj] {10.3847/1538-4357/aac2ec}, \href
  {https://ui.adsabs.harvard.edu/abs/2018ApJ...861...14C} {861, 14}

\bibitem[\protect\citeauthoryear{{Duric}}{{Duric}}{2004}]{Duric+2004book}
{Duric} N.,  2004, {Advanced astrophysics}.
Cambridge University Press, Cambridge

\bibitem[\protect\citeauthoryear{{Elia} et~al.,}{{Elia}
  et~al.}{2013}]{Elia+2013}
{Elia} D.,  et~al., 2013, \mn@doi [\apj] {10.1088/0004-637X/772/1/45}, \href
  {https://ui.adsabs.harvard.edu/abs/2013ApJ...772...45E} {772, 45}

\bibitem[\protect\citeauthoryear{{Elia} et~al.,}{{Elia}
  et~al.}{2017}]{Elia+2017}
{Elia} D.,  et~al., 2017, \mn@doi [\mnras] {10.1093/mnras/stx1357}, \href
  {https://ui.adsabs.harvard.edu/abs/2017MNRAS.471..100E} {471, 100}

\bibitem[\protect\citeauthoryear{{Elia} et~al.,}{{Elia} et~al.}{2021}]{Elia+21}
{Elia} D.,  et~al., 2021, \mn@doi [\mnras] {10.1093/mnras/stab1038}, \href
  {https://ui.adsabs.harvard.edu/abs/2021MNRAS.504.2742E} {504, 2742}

\bibitem[\protect\citeauthoryear{{Fromang}, {Hennebelle}  \&
  {Teyssier}}{{Fromang} et~al.}{2006}]{Fromang+06}
{Fromang} S.,  {Hennebelle} P.,   {Teyssier} R.,  2006, \mn@doi [\aap]
  {10.1051/0004-6361:20065371}, \href
  {http://adsabs.harvard.edu/abs/2006A%26A...457..371F} {457, 371}

\bibitem[\protect\citeauthoryear{{Fuller}, {Williams}  \& {Sridharan}}{{Fuller}
  et~al.}{2005}]{Fuller+05}
{Fuller} G.~A.,  {Williams} S.~J.,   {Sridharan} T.~K.,  2005, \mn@doi [\aap]
  {10.1051/0004-6361:20042110}, \href
  {https://ui.adsabs.harvard.edu/abs/2005A&A...442..949F} {442, 949}

\bibitem[\protect\citeauthoryear{{Giannini} et~al.,}{{Giannini}
  et~al.}{2012}]{Giannini+2012}
{Giannini} T.,  et~al., 2012, \mn@doi [\aap] {10.1051/0004-6361/201117811},
  \href {https://ui.adsabs.harvard.edu/abs/2012A&A...539A.156G} {539, A156}

\bibitem[\protect\citeauthoryear{{Gnedin} \& {Hollon}}{{Gnedin} \&
  {Hollon}}{2012}]{Gnedin+Hollon12}
{Gnedin} N.~Y.,  {Hollon} N.,  2012, \mn@doi [\apjs]
  {10.1088/0067-0049/202/2/13}, \href
  {https://ui.adsabs.harvard.edu/abs/2012ApJS..202...13G} {202, 13}

\bibitem[\protect\citeauthoryear{{Habing}}{{Habing}}{1968}]{Habing68}
{Habing} H.~J.,  1968, \bain, \href
  {http://adsabs.harvard.edu/abs/1968BAN....19..421H} {19, 421}

\bibitem[\protect\citeauthoryear{{Haugb{\o}lle}, {Padoan}  \&
  {Nordlund}}{{Haugb{\o}lle} et~al.}{2018}]{Haugboelle+2018}
{Haugb{\o}lle} T.,  {Padoan} P.,   {Nordlund} {\r{A}}.,  2018, \mn@doi [\apj]
  {10.3847/1538-4357/aaa432}, \href
  {https://ui.adsabs.harvard.edu/abs/2018ApJ...854...35H} {854, 35}

\bibitem[\protect\citeauthoryear{{Heyer} \& {Terebey}}{{Heyer} \&
  {Terebey}}{1998}]{Heyer+Terebey98}
{Heyer} M.~H.,  {Terebey} S.,  1998, \mn@doi [\apj] {10.1086/305881}, \href
  {https://ui.adsabs.harvard.edu/abs/1998ApJ...502..265H} {502, 265}

\bibitem[\protect\citeauthoryear{{Heyer}, {Brunt}, {Snell}, {Howe}, {Schloerb}
  \& {Carpenter}}{{Heyer} et~al.}{1998}]{Heyer+98}
{Heyer} M.~H.,  {Brunt} C.,  {Snell} R.~L.,  {Howe} J.~E.,  {Schloerb} F.~P.,
  {Carpenter} J.~M.,  1998, \mn@doi [\apjs] {10.1086/313086}, \href
  {http://adsabs.harvard.edu/abs/1998ApJS..115..241H} {115, 241}

\bibitem[\protect\citeauthoryear{{Juvela}}{{Juvela}}{2019}]{Juvela_2019}
{Juvela} M.,  2019, \mn@doi [\aap] {10.1051/0004-6361/201834354}, \href
  {https://ui.adsabs.harvard.edu/abs/2019A&A...622A..79J} {622, A79}

\bibitem[\protect\citeauthoryear{{Kippenhahn} \& {Weigert}}{{Kippenhahn} \&
  {Weigert}}{1994}]{Kippenhahn+1994book}
{Kippenhahn} R.,  {Weigert} A.,  1994, {Stellar Structure and Evolution}.
Springer-Verlag, Berlin Heidelberg New York

\bibitem[\protect\citeauthoryear{{Larson}}{{Larson}}{1981}]{Larson81}
{Larson} R.~B.,  1981, \mn@doi [\mnras] {10.1093/mnras/194.4.809}, \href
  {https://ui.adsabs.harvard.edu/abs/1981MNRAS.194..809L} {194, 809}

\bibitem[\protect\citeauthoryear{{Lu}, {Pelkonen}, {Padoan}, {Pan},
  {Haugb{\o}lle}  \& {Nordlund}}{{Lu} et~al.}{2020}]{Lu+20SN}
{Lu} Z.-J.,  {Pelkonen} V.-M.,  {Padoan} P.,  {Pan} L.,  {Haugb{\o}lle} T.,
  {Nordlund} {\r{A}}.,  2020, \mn@doi [\apj] {10.3847/1538-4357/abbd8f}, \href
  {https://ui.adsabs.harvard.edu/abs/2020ApJ...904...58L} {904, 58}

\bibitem[\protect\citeauthoryear{{Markwardt}}{{Markwardt}}{2009}]{Markwardt09}
{Markwardt} C.~B.,  2009, in {Bohlender} D.~A.,  {Durand} D.,   {Dowler} P.,
  eds,  Astronomical Society of the Pacific Conference Series Vol. 411,
  Astronomical Data Analysis Software and Systems XVIII. p.~251 (\mn@eprint
  {arXiv} {0902.2850})

\bibitem[\protect\citeauthoryear{{Mathis}, {Mezger}  \& {Panagia}}{{Mathis}
  et~al.}{1983}]{Mathis_1983}
{Mathis} J.~S.,  {Mezger} P.~G.,   {Panagia} N.,  1983, \aap, \href
  {https://ui.adsabs.harvard.edu/abs/1983A&A...128..212M} {500, 259}

\bibitem[\protect\citeauthoryear{{Molinari}, {Pezzuto}, {Cesaroni}, {Brand},
  {Faustini}  \& {Testi}}{{Molinari} et~al.}{2008}]{Molinari+2008}
{Molinari} S.,  {Pezzuto} S.,  {Cesaroni} R.,  {Brand} J.,  {Faustini} F.,
  {Testi} L.,  2008, \mn@doi [\aap] {10.1051/0004-6361:20078661}, \href
  {https://ui.adsabs.harvard.edu/abs/2008A&A...481..345M} {481, 345}

\bibitem[\protect\citeauthoryear{{Molinari}, {Schisano}, {Faustini},
  {Pestalozzi}, {di Giorgio}  \& {Liu}}{{Molinari} et~al.}{2011}]{Molinari+11}
{Molinari} S.,  {Schisano} E.,  {Faustini} F.,  {Pestalozzi} M.,  {di Giorgio}
  A.~M.,   {Liu} S.,  2011, \mn@doi [\aap] {10.1051/0004-6361/201014752}, \href
  {https://ui.adsabs.harvard.edu/abs/2011A&A...530A.133M} {530, A133}

\bibitem[\protect\citeauthoryear{{Molinari} et~al.,}{{Molinari}
  et~al.}{2016}]{Molinari+2016}
{Molinari} S.,  et~al., 2016, \mn@doi [\aap] {10.1051/0004-6361/201526380},
  \href {https://ui.adsabs.harvard.edu/abs/2016A&A...591A.149M} {591, A149}

\bibitem[\protect\citeauthoryear{{Ossenkopf} \& {Henning}}{{Ossenkopf} \&
  {Henning}}{1994}]{OH_1994}
{Ossenkopf} V.,  {Henning} T.,  1994, A\&A, \href
  {http://adsabs.harvard.edu/abs/1994A%26A...291..943O} {291, 943}

\bibitem[\protect\citeauthoryear{{Padoan}, {Haugb{\o}lle}  \&
  {Nordlund}}{{Padoan} et~al.}{2012}]{Padoan+12sfr}
{Padoan} P.,  {Haugb{\o}lle} T.,   {Nordlund} {\AA}.,  2012, \mn@doi [\apjl]
  {10.1088/2041-8205/759/2/L27}, \href
  {http://adsabs.harvard.edu/abs/2012ApJ...759L..27P} {759, L27}

\bibitem[\protect\citeauthoryear{{Padoan}, {Pan}, {Haugb{\o}lle}  \&
  {Nordlund}}{{Padoan} et~al.}{2016a}]{Padoan+SN1+2016ApJ}
{Padoan} P.,  {Pan} L.,  {Haugb{\o}lle} T.,   {Nordlund} {\r{A}}.,  2016a,
  \mn@doi [\apj] {10.3847/0004-637X/822/1/11}, \href
  {https://ui.adsabs.harvard.edu/abs/2016ApJ...822...11P} {822, 11}

\bibitem[\protect\citeauthoryear{{Padoan}, {Juvela}, {Pan}, {Haugb{\o}lle}  \&
  {Nordlund}}{{Padoan} et~al.}{2016b}]{Padoan+SN3+2016ApJ}
{Padoan} P.,  {Juvela} M.,  {Pan} L.,  {Haugb{\o}lle} T.,   {Nordlund}
  {\r{A}}.,  2016b, \mn@doi [\apj] {10.3847/0004-637X/826/2/140}, \href
  {https://ui.adsabs.harvard.edu/abs/2016ApJ...826..140P} {826, 140}

\bibitem[\protect\citeauthoryear{{Padoan}, {Haugb{\o}lle}, {Nordlund}  \&
  {Frimann}}{{Padoan} et~al.}{2017}]{Padoan+17sfr}
{Padoan} P.,  {Haugb{\o}lle} T.,  {Nordlund} {\r{A}}.,   {Frimann} S.,  2017,
  \mn@doi [\apj] {10.3847/1538-4357/aa6afa}, \href
  {https://ui.adsabs.harvard.edu/abs/2017ApJ...840...48P} {840, 48}

\bibitem[\protect\citeauthoryear{{Padoan}, {Pan}, {Juvela}, {Haugb{\o}lle}  \&
  {Nordlund}}{{Padoan} et~al.}{2020}]{Padoan+20massive}
{Padoan} P.,  {Pan} L.,  {Juvela} M.,  {Haugb{\o}lle} T.,   {Nordlund}
  {\r{A}}.,  2020, \mn@doi [\apj] {10.3847/1538-4357/abaa47}, \href
  {https://ui.adsabs.harvard.edu/abs/2020ApJ...900...82P} {900, 82}

\bibitem[\protect\citeauthoryear{{Pan}, {Padoan}, {Haugb{\o}lle}  \&
  {Nordlund}}{{Pan} et~al.}{2016}]{Pan+Padoan+SN2+2016ApJ}
{Pan} L.,  {Padoan} P.,  {Haugb{\o}lle} T.,   {Nordlund} {\r{A}}.,  2016,
  \mn@doi [\apj] {10.3847/0004-637X/825/1/30}, \href
  {https://ui.adsabs.harvard.edu/abs/2016ApJ...825...30P} {825, 30}

\bibitem[\protect\citeauthoryear{{Peretto} et~al.,}{{Peretto}
  et~al.}{2013}]{Peretto+13}
{Peretto} N.,  et~al., 2013, \mn@doi [\aap] {10.1051/0004-6361/201321318},
  \href {https://ui.adsabs.harvard.edu/abs/2013A&A...555A.112P} {555, A112}

\bibitem[\protect\citeauthoryear{{Salaris} \& {Cassisi}}{{Salaris} \&
  {Cassisi}}{2005}]{Salaris+2005book}
{Salaris} M.,  {Cassisi} S.,  2005, {Evolution of Stars and Stellar
  Populations}.
Wiley-VCH, Weinheim

\bibitem[\protect\citeauthoryear{{Salpeter}}{{Salpeter}}{1955}]{Salpeter55}
{Salpeter} E.~E.,  1955, \mn@doi [\apj] {10.1086/145971}, \href
  {https://ui.adsabs.harvard.edu/abs/1955ApJ...121..161S} {121, 161}

\bibitem[\protect\citeauthoryear{{Schaller}, {Schaerer}, {Meynet}  \&
  {Maeder}}{{Schaller} et~al.}{1992}]{Schaller+92}
{Schaller} G.,  {Schaerer} D.,  {Meynet} G.,   {Maeder} A.,  1992, \aaps, \href
  {http://adsabs.harvard.edu/abs/1992A%26AS...96..269S} {96, 269}

\bibitem[\protect\citeauthoryear{{Tan}, {Beltr{\'a}n}, {Caselli}, {Fontani},
  {Fuente}, {Krumholz}, {McKee}  \& {Stolte}}{{Tan} et~al.}{2014}]{Tan+2014}
{Tan} J.~C.,  {Beltr{\'a}n} M.~T.,  {Caselli} P.,  {Fontani} F.,  {Fuente} A.,
  {Krumholz} M.~R.,  {McKee} C.~F.,   {Stolte} A.,  2014, in {Beuther} H.,
  {Klessen} R.~S.,  {Dullemond} C.~P.,   {Henning} T.,  eds, Protostars and
  Planets VI. p.~149 (\mn@eprint {arXiv} {1402.0919}),
  \mn@doi{10.2458/azu\_uapress\_9780816531240-ch007}

\bibitem[\protect\citeauthoryear{{Teyssier}}{{Teyssier}}{2002}]{Teyssier+2002A&A}
{Teyssier} R.,  2002, \mn@doi [\aap] {10.1051/0004-6361:20011817}, \href
  {https://ui.adsabs.harvard.edu/abs/2002A&A...385..337T} {385, 337}

\bibitem[\protect\citeauthoryear{{Teyssier}}{{Teyssier}}{2007}]{Teyssier07}
{Teyssier} R.,  2007, \mn@doi [Geophysical and Astrophysical Fluid Dynamics]
  {10.1080/03091920701523386}, \href
  {http://adsabs.harvard.edu/abs/2007GApFD.101..199T} {101, 199}

\bibitem[\protect\citeauthoryear{{Traficante}, {Fuller}, {Billot},
  {Duarte-Cabral}, {Merello}, {Molinari}, {Peretto}  \&
  {Schisano}}{{Traficante} et~al.}{2017}]{Traficante+17}
{Traficante} A.,  {Fuller} G.~A.,  {Billot} N.,  {Duarte-Cabral} A.,  {Merello}
  M.,  {Molinari} S.,  {Peretto} N.,   {Schisano} E.,  2017, \mn@doi [\mnras]
  {10.1093/mnras/stx1375}, \href
  {https://ui.adsabs.harvard.edu/abs/2017MNRAS.470.3882T} {470, 3882}

\bibitem[\protect\citeauthoryear{{Traficante} et~al.,}{{Traficante}
  et~al.}{2018}]{Traficante+18}
{Traficante} A.,  et~al., 2018, \mn@doi [\mnras] {10.1093/mnras/sty798}, \href
  {https://ui.adsabs.harvard.edu/abs/2018MNRAS.477.2220T} {477, 2220}

\bibitem[\protect\citeauthoryear{{Urquhart} et~al.,}{{Urquhart}
  et~al.}{2014}]{Urquhart+2014}
{Urquhart} J.~S.,  et~al., 2014, \mn@doi [\mnras] {10.1093/mnras/stu1207},
  \href {https://ui.adsabs.harvard.edu/abs/2014MNRAS.443.1555U} {443, 1555}

\bibitem[\protect\citeauthoryear{{Wolfire}, {Hollenbach}, {McKee}, {Tielens}
  \& {Bakes}}{{Wolfire} et~al.}{1995}]{Wolfire+95}
{Wolfire} M.~G.,  {Hollenbach} D.,  {McKee} C.~F.,  {Tielens} A.~G.~G.~M.,
  {Bakes} E.~L.~O.,  1995, \mn@doi [\apj] {10.1086/175510}, \href
  {http://adsabs.harvard.edu/abs/1995ApJ...443..152W} {443, 152}

\bibitem[\protect\citeauthoryear{{Wyrowski} et~al.,}{{Wyrowski}
  et~al.}{2016}]{Wyrowski+16}
{Wyrowski} F.,  et~al., 2016, \mn@doi [\aap] {10.1051/0004-6361/201526361},
  \href {https://ui.adsabs.harvard.edu/abs/2016A&A...585A.149W} {585, A149}

\bibitem[\protect\citeauthoryear{{Yuan} et~al.,}{{Yuan} et~al.}{2018}]{Yuan+18}
{Yuan} J.,  et~al., 2018, \mn@doi [\apj] {10.3847/1538-4357/aa9d40}, \href
  {https://ui.adsabs.harvard.edu/abs/2018ApJ...852...12Y} {852, 12}

\makeatother
\end{thebibliography}

\begin{appendix}

\section{Description of the synthetic catalog} \label{appA}

We generate the synthetic catalog that used to study the real nature of massive clumps in the galaxy in this paper, defined as follows:

\begin{itemize}

\item Column[1], $ID$: running number (starting from 1 to 52,543).
\item Column[2], $SNAPSHOT$: the number of snapshots (839, 1107 and 1479, respectively).
\item Column[3], $DIRECTION$: the direction of the maps (0, 1, 2 for x, y and z directions, respectively).
\item Column[4], $DISTANCE$: the distance of the sources, in pc.
\item Column[5], $X250$, [6], $Y250$: the map coordinates of the sources detected at 250 $\mu$m, in box units (from 0 to 1). Column [7], $Z250$: the line-of-sight coordinate of the maximum density in the 3D density cube, in box units (from 0 to 1).
\item Column[8], $FWHM\_X$, and [9], $FWHM\_Y$: the Full Width Half Maximum of the fitted bi-dimensional Gaussian along the x-axis and y-axis, in box units (from 0 to 1).
\item Column[10], $ANGLE$: the orientation angle of the major axis of the Gaussian respect to the x-axis, in degree.
\item Column[11], $R$: the deconvolved radius of the source estimated at 250 $\mu$m, in pc.
\item Column[12], $F70$, [13], $DF70$, [14], $F160$, [15], $DF160$, [16], $F250$, [17], $DF250$, [18], $F350$, [19], $DF350$, [20], $F500$ and [21], $DF500$: the integrated flux of the source, in Jy, and noise of the flux, in Jy, estimated from the RMS for 70, 160, 250, 350 and 500 $\mu$m. Null value is -1.
\item Column[22], $LAMBDA\_0$: the wavelength $\lambda_{0}$, in $\mu$m, from SED fitting for optical thick cases, and 0 for the optical thin cases.
\item Column[23], $T$, and [24], $DT$: temperature and its error from SED fitting, in K.
\item Column[25], $M\_SED$, and [26], $DM\_SED$: the mass of the clump and its error calculated from SED fitting with a greybody function, in $\rm M_{\odot}$.
\item Column[27], $F70\_GB$: the flux at 70 $\mu$m calculated from SED fitting with greybody function, in Jy.
\item Column[28], $TYPE$: the classification of clumps: starless unbound, prestellar and protostellar are denoted as 0, 1 and 2, respectively.
\item Column[29], $M\_MAIN$: the mass of the main 3D clump along the light of sight, in $\rm M_{\odot}$.
\item Column[30], $M\_ALL$: the mass of all the 3D clumps with density peaks $\geq1\%$ of the maximum density peak along the light of sight, in $\rm M_{\odot}$.
\item Column[31], $M\_TOT$: the total mass of column along the light of sight, in $\rm M_{\odot}$.
\item Column[32], $LOCAL\_GAS\_DENSITY$: the maximum of local gas densities around stars along the line of sight, in $\rm cm^{-3}$. Null value is 0.
\item Column[33], $STAR\_AGE$: the maximum age of the stars along the line of sight, in Myr. Null value is 0.
\item Column[34], $STAR\_MASS$: the maximum mass of the stars along the line of sight, in $\rm M_{\odot}$. Null value is 0.

\end{itemize}

\end{appendix}

\bsp	
\label{lastpage}
\end{document}